\documentstyle[prc,aps,preprint,psfig]{revtex}

\draft
\begin{document}

\title{Open charm and charmonium production at relativistic energies}

\author{E.~L.~Bratkovskaya$^1$\thanks{Supported by DFG},
W. Cassing$^2$ and H.~St\"ocker$^{1}$\\[5mm]
{$^1$ \normalsize Institut f\"{u}r Theoretische Physik,
    Universit\"{a}t Frankfurt}\\
    {\normalsize 60054 Frankfurt, Germany}\\[3mm]
{$^2$ \normalsize Institut f\"{u}r Theoretische
    Physik, Universit\"{a}t Giessen}\\
    {\normalsize 35392 Giessen, Germany}}
\maketitle

\begin{abstract}
We calculate open charm and charmonium production in $Au+Au$
reactions at $\sqrt{s}$ = 200 GeV within the hadron-string
dynamics (HSD) transport approach employing open charm cross
sections from $pN$ and $\pi N$ reactions that are fitted to
results from PYTHIA and scaled in magnitude to the available
experimental data. Charmonium dissociation with nucleons and
formed mesons to open charm ($D+\bar{D}$ pairs) is included
dynamically. The 'comover' dissociation cross sections are
described by a simple phase-space model including a single free
parameter, i.e. an interaction strength $M_0^2$, that is fitted to
the $J/\Psi$ suppression data for $Pb+Pb$ collisions at SPS
energies.  As a novel feature we implement the backward channels
for charmonium reproduction by $D \bar{D}$ channels employing
detailed balance. From our dynamical calculations we find that the
charmonium recreation is comparable to the dissociation by
'comoving' mesons. This leads to the final result that the total
$J/\Psi$ suppression at $\sqrt{s}$ = 200 GeV as a function of
centrality is slightly less than the suppression seen at SPS
energies by the NA50 Collaboration, where the 'comover'
dissociation is substantial and the backward channels play no
role. Furthermore, even in case that all directly produced
$J/\Psi$ mesons dissociate immediately (or are not formed as a
mesonic state), a sizeable amount of charmonia is found
asymptotically due to the $D+\bar{D} \rightarrow J/\Psi$ + meson
channels in central collisions of $Au+Au$ at $\sqrt{s}$ = 200 GeV
which, however, is lower than the $J/\Psi$ yield expected from
binary scaling of $pp$ collisions.
\end{abstract}

\noindent
\vspace{10mm} \noindent



\newpage
\narrowtext

\section{Introduction}
The dynamics of ultra-relativistic nucleus-nucleus collisions at SPS
and RHIC energies are of fundamental interest with respect to the
properties of hadronic/partonic systems at high energy densities as
encountered in the early phase of the 'big bang'.  Especially the
formation of a quark-gluon plasma (QGP) and its transition to
interacting hadronic matter has motivated a large community for about
20 to 30 years by now \cite{QM01}. However, even after more than a
decade of experiments at the Super Proton Synchrotron (SPS) and
 recently at the Relativistic Heavy-Ion Collider (RHIC) the complexity
of the dynamics has not been unraveled and no conclusive evidence has
been obtained for the formation of the QGP and/or the properties of the
phase transition \cite{Bass99,Reiter} though 'circumstantial evidence'
has been claimed \cite{Heinz}.

Apart from the light and strange flavor
($u,\bar{u},d,\bar{d},s,\bar{s}$) quark physics and their hadronic
bound states in the vacuum ($\pi, K, \phi$ etc.) the interest in
hadronic states with charm flavors ($c, \bar{c}$) has been rising
additionally in line with the development of new experimental
facilities. This relates to the charm production cross section in $pN$,
$\pi N$, $pA$ and $AA$  reactions as well as to their interactions with
baryons and mesons which determine their properties (spectral
functions) in the hadronic medium.

The charm quark degrees of freedom are of special interest in context
with the phase transition to the  QGP since $c\bar{c}$ meson states
should no longer be formed due to color screening \cite{Satz,Satznew}.
However, the suppression of $J/\Psi$ and $\Psi^\prime$ mesons in the
high density phase of nucleus-nucleus collisions at SPS energies
\cite{NA50aa,NA50bb,NA50b,NA50a,NA50_QM02} might also be attributed to
inelastic comover scattering (cf.
\cite{Capella,Cass99,Vogt99,Gersch,Cass00,Spieles,Spieles2,Gerland} and
Refs. therein) provided that the corresponding $J/\Psi$-hadron cross
sections are in the order of a few mb
\cite{Haglin,Haglin2,Konew,Ko,Sascha,Sascha2,wong02,david}.
Theoretical estimates here differ by more than an order of magnitude
\cite{Bernd} especially with respect to $J/\Psi$-meson scattering such
that the question of charmonium suppression is not yet settled. On the
other hand, at RHIC energies further absorption mechanisms -- such as
plasma screening and gluon scattering -- might play a dominant role as
suggested in Refs.  \cite{Kojpsi,Rappnew} and also lead to a
substantial reduction of the $J/\Psi$ formation  in central $Au+Au$
collisions.

On the other hand, it has been pointed out -- within statistical models
-- that at RHIC energies the charmonium formation from open charm +
anticharm mesons might become essential \cite{Rafelski} and even exceed
the yield from primary $NN$ collisions \cite{Rafelski,Johanna}.
However, a more schematic model by Ko et al. \cite{CMKO} -- including
the channels $J/\Psi + \pi \leftrightarrow D \bar{D}$ suggested that
such channels should be still of minor importance at RHIC energies but
become essential at LHC energies. A similar conclusion has been reached
in Ref.  \cite{Redlich}. One of the prevailing questions thus is, if
open charm mesons and charmonia will achieve thermal and chemical
equilibrium with the light mesons during the nucleus-nucleus reaction
as suggested/anticipated in Refs. \cite{Marek1,Marek2,Gall,Go1}.  Such
issues of equilibration phenomena are traditionally examined within
nonequilibrium relativistic transport theory
\cite{Cass99,Horst86,Cass90,Koreview,UrQMD1}.

In this work we will calculate open charm and charmonium production at
RHIC energies within the HSD transport approach
\cite{Cass99,Cass00,Cass01} for the overall reaction dynamics using
parametrizations for the elementary production channels including the
charmed hadrons $D, \bar{D}, D^*, \bar{D}^*, D_s, \bar{D}_s, D_s^*,
\bar{D}_s^*,$ $J/\Psi, \Psi(2S), \chi_{2c}$ from $NN$ and $\pi N$
collisions. The latter parametrizations are fitted to PYTHIA
calculations \cite{PYTHIA} above $\sqrt{s}$ = 10 GeV and extrapolated
to the individual thresholds, while the absolute strength of the cross
sections is fixed by the experimental data as described in Ref.
\cite{Cass01}. In the latter work we have calculated excitations
functions for open charm mesons and charmonia including the $J/\Psi$
suppression by dissociation with baryons and meson ('comovers')  using
the $J/\Psi$-meson cross sections from Haglin \cite{Haglin}. The
centrality dependence for the $J/\Psi$ survival probability has been
presented in Ref. \cite{Cass00} for SPS ($\sqrt{s}$ = 17.3 GeV) and
RHIC energies ($\sqrt{s}$ = 200 GeV), too, for $Pb+Pb$ or $Au+Au$
collisions, respectively. We here extend our previous works and include
explicitly the backward channels 'charm + anticharm meson $\rightarrow$
charmonia + meson' employing detailed balance in a more schematic
interaction model with a single parameter or matrix element $|M_0|^2$,
that is fixed by the $J/\Psi$ suppression data from the NA50
collaboration at SPS energies.

Our work is organized as follows: In Section 2 we will present the
results of the HSD transport approach for charged hadrons, protons,
antiprotons and elliptic flow in $Au+Au$ collisions at $\sqrt{s}$ = 200
GeV in comparison to available data. This presentation is necessary
since the open and hidden charm formation and propagation proceeds in a
dense and hot hadronic environment that should be sufficiently
realistic. The elementary production cross sections for open charm and
charmonia  from baryon-baryon ($BB$) and meson-baryon ($mB$) collisions
are presented in Section 3 as well as their interaction cross sections
with hadrons. A phase-space model will be presented, furthermore, for
the charmonium + meson dissociation cross sections that allows to
implement 'detailed balance' for all channels of interest.  Section 4
contains the actual calculations for the open and hidden charm degrees
of freedom for $Pb+Pb$ collisions at $\sqrt{s}$ = 17.3 GeV and $Au+Au$
collisions at $\sqrt{s}$ = 200 GeV with particular emphasis on the
novel aspect, i.e. the charmonium reformation by open charm mesons
employing 'detailed balance'. A comparison to the preliminary data of
the PHENIX Collaboration on $J/\Psi$ suppression in $Au+Au$ collisions
at $\sqrt{s}$ = 200 GeV will be presented, too. Section 5 concludes
this study with a summary and discussion of open problems.

\section{Charged hadrons, baryons, antibaryons and collective flow}

Before coming to the actual charmonium and open charm dynamics at RHIC
energies we have to investigate, if the HSD transport approach based on
string, quark, diquark ($q, \bar{q}, qq, \bar{q}\bar{q}$) as well as
hadronic degrees of freedom performs reasonably well with respect to
the abundancy of light hadrons composed of $u,d,s$ quarks\footnote{For
a more recent survey on hadron rapidity distributions from 2 to 160
A~GeV in central nucleus-nucleus collisions within the HSD and UrQMD
\cite{Weber} transport approaches we refer the reader to Ref.
\cite{brat02}.}.  Such a test is essential since the dissociation of
charmonia on baryon, antibaryons and mesons is directly proportional to
their density in phase space. We recall that in HSD all newly produced
hadrons have a formation time of $\tau_F$ = 0.8 fm/c in their rest
frame and do not interact during the 'partonic' propagation.
Furthermore, hadronization is inhibited if the energy density -- in the
local rest frame -- is above 1 GeV/fm$^3$, which roughly corresponds to
the energy density for QGP formation in equilibrium at vanishing quark
chemical potential $\mu_q$. Thus 'hadrons' only exist as
quark-antiquark or quark-diquark pairs at energy densities above 1
GeV/fm$^3$ and only can become ordinary hadrons if the system has
expanded sufficiently. We note that this cut on the energy density is
the only modification introduced as compared to the earlier studies in
Refs.  \cite{Cass00,Cass01,Cass02} and has also been included in the
more recent systematic analysis in Ref. \cite{brat02} from SIS to SPS
energies.

\begin{figure}[h]
\centerline{\psfig{file=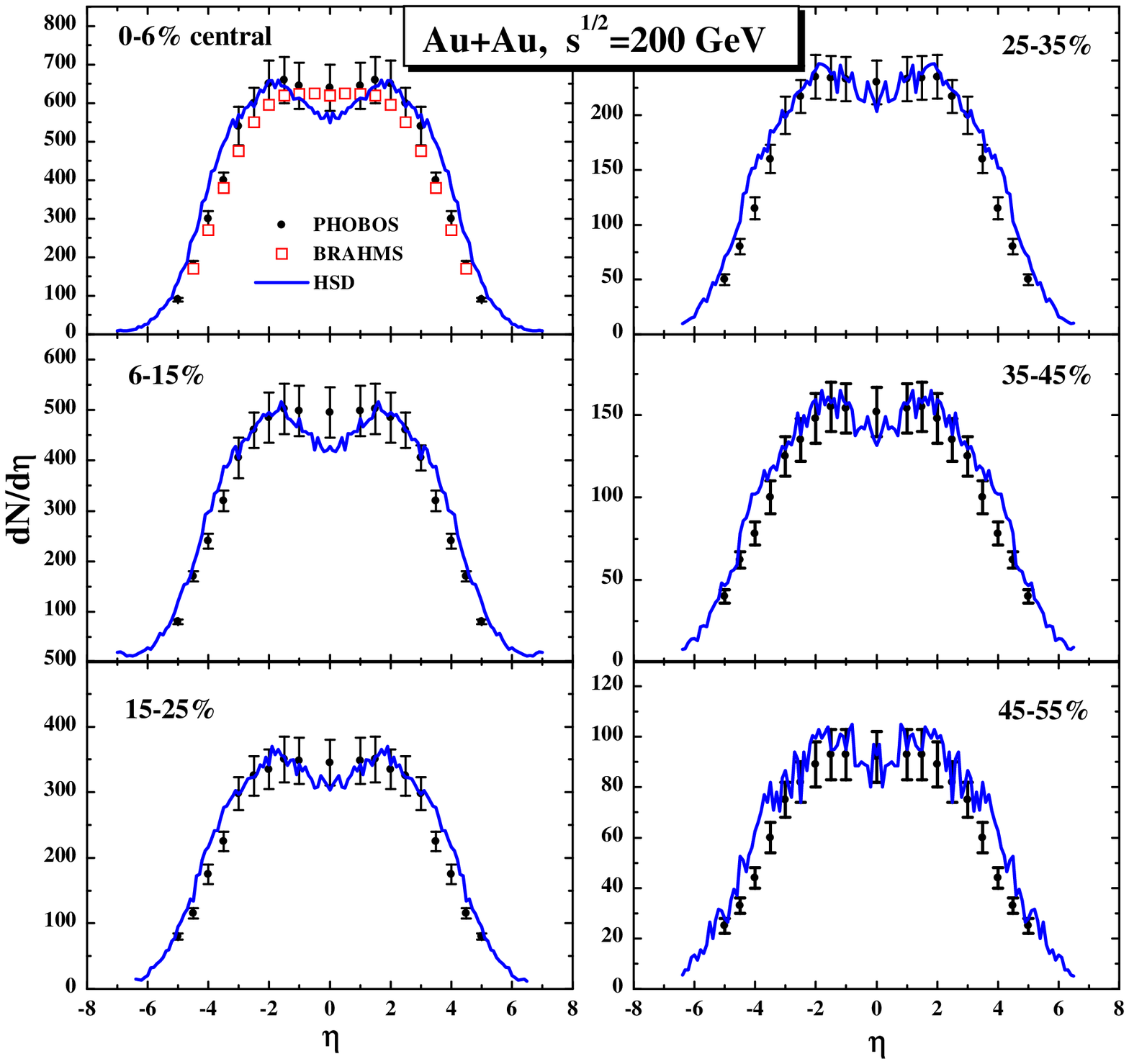,width=16cm}}
\vspace*{5mm}
\caption{The calculated pseudo-rapidity distributions of charged
hadrons (solid lines) for $Au+Au$ at $\sqrt{s}$ = 200 GeV for different
centrality classes in comparison to the experimental data of the PHOBOS
Collaboration \protect\cite{PHOBOS} (full points), where the error bars
indicate the systematic experimental uncertainty. The open squares in
the upper left figure correspond to the data from the BRAHMS
Collaboration for the same centrality class \protect\cite{BRAHMS}.  }
\label{bild1} \end{figure}

In order to demonstrate the applicability of the HSD approach to
nucleus-nucleus collisions at RHIC energies we show in Fig. \ref{bild1}
the calculated pseudo-rapidity distributions of charged hadrons (solid
lines) for $Au+Au$ at $\sqrt{s}$ = 200 GeV for different centrality
classes in comparison to the experimental data of the PHOBOS
Collaboration \cite{PHOBOS} (full points), where the error bars
indicate the systematic experimental uncertainty. The open squares in
the upper left figure correspond to the data from the BRAHMS
Collaboration for the same centrality class \cite{BRAHMS}. We find that
the HSD calculations show a small dip in $dN/d\eta$ at midrapidity for
all centrality classes, which is not seen in the experimental
distributions. Furthermore, the pseudo-rapidity distributions are
slightly broader than the data which also might point towards an
improper string fragmentation scheme in the LUND model \cite{LUND}
employed in HSD. We expect that this issue can be settled uniquely when
high statistics data for $pp$ reactions at RHIC energies become
available. On the other hand, the overall description of the rapidity
distributions is reasonable good for our present purposes.

A further question is related to the antibaryon and baryon abundancies
at midrapidity that show the amount of baryon stopping and antibaryon
production \cite{Weber02}. We mention that multi-meson fusion channels
play a sizeable role in recreating baryon-antibaryon pairs
\cite{Cass02,Rapp,Greiner} and reducing the number of light mesons
accordingly. Thus detailed balance on the many-particle level -- as
only found more recently \cite{Cass02} -- leads to an approximate
chemical equilibrium of antibaryons with mesons whenever the meson
density is sufficiently high as e.g. in nonperipheral $Au+Au$
collisions at RHIC energies.  Our numerical results for the $(\bar{p} +
\bar{\Lambda})/(p + \Lambda)$ ratio in 10\% central $Au+Au$ collisions
at $\sqrt{s}$ = 200 GeV are displayed in Fig. \ref{bild2} as a function
of rapidity $y$ in comparison to the data from the BRAHMS Collaboration
\cite{BRAHMS2}, that correspond to the measured $\bar{p}/p$ ratio,
however, include some still unknown fraction from $\Lambda$ and ${\bar
\Lambda}$ decays.  The comparison in Fig. \ref{bild2} thus suffers from
a 5--10\% systematic uncertainty. We mention that (within statistics)
practically the same rapidity distribution for antiprotons is obtained
when discarding baryon-antibaryon annihilation as well as the backward
channels. Thus the calculations for charmonia and open charm mesons in
Section 4 will be performed in the latter limit. Nevertheless, Fig.
\ref{bild2} suggests that the antiproton/proton ratio is reasonably
described in the HSD approach. This also holds for the net proton
$(p-\bar{p})$ rapidity distribution as seen from Fig. \ref{bild3} in
comparison to the preliminary data of the BRAHMS Collaboration
\cite{BRAHMS3} for the same event class as in Fig.
\ref{bild2}\footnote{The experimental data again include some unknown
fraction of $\Lambda$ and $\bar{\Lambda}$ decays such that the 'real'
$(p-\bar{p})$ rapidity distribution should be slightly lower.}.

In principle, one might argue that a transport approach based on string
and hadronic degrees of freedom should not be adequate in the initial
stage of nucleus-nucleus collisions at RHIC energies where a new state
of matter, i.e. a quark-gluon plasma (QGP), is expected/hoped to be
formed. However, the global event characteristics and particle
abundancies from SIS to RHIC energies are found experimentally to show
a rather smooth evolution with bombarding energy \cite{survey,Baker}
such that no obvious conclusion on the effective degrees of freedom in
the initial phase can presently be drawn. Moreover, the large pressure
needed to describe the elliptic flow at RHIC energies is approximately
described by 'early' hadron formation -- as in HSD -- and the 'large'
hadronic interaction cross sections. This is demonstrated in Fig.
\ref{bild4} where we show the calculated elliptic flow $v_2$ for
charged hadrons (solid lines) as a function of the pseudorapidity
$\eta$ (upper part) and as a function of the number of 'participating
nucleons' $N_{part}$ (lower part)  for $|\eta| \leq$ 1 in comparison to
the preliminary 'hit-based analysis' data of the PHOBOS Collaboration
\cite{PHOBOS1}.  Note, that the experimental error bars correspond to
1$\sigma$ statistical errors, only. Our calculations underestimate  the
$v_2(\eta)$ distribution close to midrapidity and also are somewhat low
in the centrality dependence of the elliptic flow. Whereas the elliptic
flow at midrapidity is well described by hydrodynamical models, the
$v_2(\eta)$ distribution comes out too flat in these calculations
\cite{Hirano}. We note, that our HSD results are very similar to those
of the hadronic rescattering model by Humanic et al. \cite{Tom1,Tom2}
and almost quantitatively agree with the calculations by Sahu et al.
\cite{Sahu02} performed within the hadron-string cascade model JAM
\cite{JAM}.

On the other hand, unexpectedly high parton cross sections of $\sim$
5--6 mb have to be assumed in parton cascades \cite{Texas} in order to
reproduce the elliptic flow $v_2(p_T)$ seen experimentally.  These
cross sections are about 1/9 of the baryon-baryon total cross section
($\sim$ 45 mb) or 1/6 of the meson-baryon cross section ($\sim$ 30 mb)
such that the effective cross section for the constituent quarks and
antiquarks is roughly the same in the partonic and hadronic phase. In
this context it will be important to have precise data on open charm
and charmonium transverse momentum ($p_T$) spectra since their slope
might give information on the pressure generated in a possible partonic
phase \cite{Xu}. This argument is expected to hold especially for
$J/\Psi$ mesons since their elastic rescattering cross section with
hadrons should be small in the hadronic expansion phase \cite{Heinz95}.
We note, that in central $Pb+Pb$ collisions at SPS energies the
spectral slope of $J/\Psi$ mesons is found experimentally to be
substantially smaller ($\sim$ 240 MeV \cite{NA50c}) than that of
protons ($\sim$ 300 MeV \cite{Bearden}). At RHIC energies the radial
flow in central $Au+Au$ collisions is even larger leading to a stiffer
spectrum with an inverse slope parameter $\sim$ 400 MeV for the
strongly interacting protons \cite{STAR}.

Nevertheless, in addition to nucleus-nucleus collisions from SIS to SPS
energies \cite{brat02}  the HSD transport approach is found to work
reasonably well also at RHIC energies for the 'soft' hadron abundancies
such that the 'hadronic environment' for open charm mesons and
charmonia should be sufficiently realistic.

\section{Elementary cross sections}

In order to examine the dynamics of open charm and charmonium degrees
of freedom during the formation and expansion phase of the highly
excited system one has to know the number of initially produced
particles with $c$ or $\bar{c}$ quarks, i.e. $D, \bar{D}, D^*,
\bar{D}^*, D_s, \bar{D}_s, D_s^*, \bar{D}_s^*,$ $J/\Psi, \Psi(2S),
\chi_{2c}$.

\subsection{Production cross section is $pp$ and $\pi N$
collisions}

In Ref. \cite{Cass01} we have fitted the total
charmonium cross sections ($X = \chi_C,  J/\Psi, \Psi^\prime$) from
$NN$ collisions as a function of the invariant energy $\sqrt{s}$
by the function
\begin{equation}
\sigma_X^{NN} (s) = b_X \left(1 - {m_X\over \sqrt{s}}\right)^\alpha
\ \left({m_X\over \sqrt{s}}\right)^{-\beta} \ \Theta(\sqrt{s}-\sqrt{s_0})
 \label{fitj}
\end{equation}
with $\alpha$ = 10, $\beta =1$, while $\sqrt{s_0}$ denotes the
threshold in vacuum. The parameters were fixed in \cite{Cass01} to
describe the  $J/\Psi$ and $\Psi^\prime$ data at lower energy
($\sqrt{s} \leq$ 30 GeV).  For our present study we use the same
parametrization (\ref{fitj}) with a slightly modified parameter
$\beta=0.775$ (instead of $\beta=1$) in order to fit the preliminary
data point from the PHENIX Collaboration \cite{PHENIX} at
$\sqrt{s}=200$ GeV, which gives $\sigma(pp\to J/\Psi +X) = 3.8\pm
0.6(\rm{stat.})\pm 1.3(\rm{sys.})$ $\mu$b for the total $J/\Psi$ cross
section. The parameter $b_X = 240 \ C_X$~nb is proportional to the
fraction of charmonium states $C_X$. We choose $C_{\chi_C}=0.4, \
C_{J/\Psi}=0.46,\ C_{\Psi^\prime}=0.14$ in line with Ref. \cite{Gavai}.

For the total charmonium cross sections from $\pi N$ reactions
we adopt the parametrization
(in line with Ref. \cite{Vogt99}):
\begin{eqnarray}
\sigma_X^{\pi N} (s) = d_X \left(1 - {m_X\over \sqrt{s}}\right)^\gamma
\label{fitpin}\end{eqnarray}
with $\gamma=7.3$ and $d_x=1360.8 \ C_X$~nb, which describes the
existing experimental data at low $\sqrt{s}$ reasonably well
(cf. Fig. 3 from \cite{Cass01}).

Apart from the total cross sections, we also need the differential
distribution of the produced mesons in the transverse momentum $p_T$
and the rapidity $y$ (or Feynman $x_F$) from each individual collision.
We recall that $x_F = p_z/p_z^{max} \approx 2 p_z/\sqrt{s}$ with $p_z$
denoting the longitudinal momentum. For the differential distribution
in $x_F$ from $NN$ and $\pi N$ collisions we use the ansatz from
the E672/E706 Collaboration \cite{E672}:
\begin{equation}
\frac{dN}{dx_F dp_T} \sim (1 - |x_F|)^c \ \exp(-b_{p_T} p_T),
\label{fit2}
\end{equation}
where $b_{p_T}=2.08$ GeV$^{-1}$ and $c= a/(1+b/\sqrt{s})$.
The parameters $a, b$ are choosen as $a_{NN}=13.5$, $b_{NN}=24.9$ for
$NN$ collisions and $a_{\pi N}=4.11$, $b_{\pi N}=10.2$ for $\pi N$
collisions.

In Fig. \ref{bild5} (upper part) we compare the calculated $J/\Psi$
differential cross section in rapidity $y_{cm}$ -- multiplied by the
branching ratio to dileptons -- with the preliminary data from the
PHENIX Collaboration \cite{PHENIX} for $pp$ collisions at $\sqrt{s}$ =
200 GeV using $\beta = 0.775$. Our elementary $J/\Psi$ formation is
seen to be in sufficient agreement with the preliminary data
\cite{PHENIX} though the rapidity distribution appears slightly broader
than the data.

The number of primary $J/\Psi$ mesons formed in central $Au+Au$
reactions at $\sqrt{s}$ = 200 GeV can be estimated -- on the basis of
the Glauber model -- by multiplying the $pp$ production cross section
with the number of binary collisions ($N_{bin} \approx 1.2\times10^3$)
and dividing by the inelastic $pp$ cross section ($\sim 45$ mb).  This
leads to a multiplicity of primary $J/\Psi$'s of $\sim 0.1$ in very
central $Au+Au$ collisions.

The total and differential cross sections for open charm mesons from
$pp$ collisions, furthermore, are taken as in Ref.  \cite{Cass01}. They
also might have to be reduced slightly as the charmonia cross sections,
however, no experimental constraint is available so far. We thus refer
to the results of Ref.  \cite{Cass01} which give $\sim$16 $D\bar{D}$
pairs in central $Au+Au$ collisions at $\sqrt{s}$ = 200 GeV, a factor
of $\sim$160 relative to the expected primordial $J/\Psi$ multiplicity
of $\sim 0.1$.  Note, that at $\sqrt{s} \approx$ 17.3 GeV the primary
$D\bar{D}$ to $J/\Psi$ ratio is about 40 \cite{Cass01}; the increase of
this ratio by a factor of $\sim$ 4 from $\sqrt{s}$ = 17.3 GeV to
$\sqrt{s}$ = 200 GeV is within the expected range. Our results for the
rapidity distribution of open charm mesons from $pp$ collisions at
$\sqrt{s}$ = 200 GeV (summing up all $D$ and $\bar{D}$ mesons) is
displayed in the lower part of Fig. \ref{bild5} and shows a rather flat
distribution at midrapidity, too. Presently, there are no data that
could control this open charm rapidity spectrum.

Apart from primary hard $NN$ collisions the open charm mesons or
charmonia may also be generated by secondary 'meson'-'baryon' ($mB$)
reactions. Here we include all secondary collisions of mesons with
'baryons' by assuming that the open charm cross section (from Section 2
of Ref. \cite{Cass01}) only depends on the invariant energy $\sqrt{s}$
and not on the explicit meson or baryon state.  Furthermore, we take
into account all interactions of 'formed' mesons -- after a formation
time of $\tau_F$ = 0.8 fm/c (in their rest frame) \cite{Geiss} -- with
baryons or diquarks, respectively. As pointed out in Ref. \cite{Cass01}
the production of open charm pairs in central $Au+Au$ collisions by
$mB$ reactions is expected to be on the 10\% level.

In order to study the effect of rescattering we tentatively adopt
the following dissociation cross sections of charmonia with
baryons independent on the energy (in line with Refs. \cite{Cass00,Cass01}):
\begin{equation}
\sigma_{c\bar{c}B} = 6 \ {\rm mb}; \ \sigma_{J/\Psi B} = 4 \ {\rm
mb}; \ \sigma_{\chi_c B} = 5 \ {\rm mb}; \  \sigma_{\Psi^\prime B}
= 10 \ {\rm mb}. \label{sigmacB}
\end{equation}
In (\ref{sigmacB}) the cross section $\sigma_{c\bar{c}B}$ stands for a
(color dipole)  pre-resonance ($c\bar{c})$ - baryon cross section,
since the $c\bar{c}$ pair produced initially cannot be identified with
a particular hadron due to the uncertainty relation in energy and time.
For the lifetime of the pre-resonance $c\bar{c}$ pair (in it's rest
frame) a value of $\tau_{c\bar{c}}$ = 0.3 fm/c is assumed following
Ref. \cite{Kharz}. This value corresponds to the mass difference of the
$\Psi^\prime$ and $J/\Psi$.

For $D, D^*, \bar{D}, \bar{D}^*$ - meson ($\pi, \eta, \rho, \omega$)
scattering we address to the calculations from Ref.  \cite{Konew,Ko}
which predict elastic cross sections in the range of 10--20 mb
depending on the size of the formfactor employed. As a guideline we use
a constant cross section of 10 mb for elastic scattering with mesons
and also baryons, although the latter might be even higher for very low
relative momenta.

\subsection{Comover dissociation channels}

As already pointed out in the introduction the $J/\Psi$ formation
cross sections by open charm mesons or the inverse comover
dissociation cross sections are not well known and the
significance of these channels is discussed controversely in the
present literature \cite{Bernd,Rafelski,Johanna,Redlich,I2,I3}. Whereas
in Refs. \cite{Cass00,Cass01} the energy-dependent $J/\Psi$-meson
cross sections for dissociation to $D\bar{D}$ have been taken from
the calculations of Haglin \cite{Haglin}, we here introduce a
simple 2-body transition model with a single free parameter
$M_0^2$, that allows to implement the backward reactions uniquely
by employing detailed balance for each individual channel.
Since the meson-meson dissociation
and backward reactions typically occur with low relative momenta
('comovers') it is legitimate to write the cross section for the
process $m_1 + m_2 \rightarrow m_3 +m_4$ as
\begin{equation}
\label{model}
 \sigma_{1+2 \rightarrow 3+4}(\sqrt{s}) = 2^4
 \ \frac{E_1 E_2 E_3
E_4}{s} \ |M_f|^2 \ \left( \frac{M_3+M_4}{\sqrt{s}} \right)^6 \
\frac{P_f}{P_i},
\end{equation}
 where $E_i$ and $S_i$ denote the energy and spin of hadron $i$,
respectively. The initial and final momenta for fixed invariant
energy  $\sqrt{s}$ are given by
\begin{equation}
P_i^2 = \frac{(s-(M_1+M_2)^2)(s-(M_1-M_2)^2)}{4s}, \ P_f^2 =
\frac{(s-(M_3+M_4)^2)(s-(M_3-M_4)^2)}{4s}, \label{moment}
\end{equation}
where $M_i$ denotes the mass of hadron $i$. In
(\ref{model}) $|M_f|^2$ stands for the effective matrix element
squared  which for the different 2-body channels is taken of the
form
\begin{eqnarray}
& |M_f|^2 = M_0^2 \hspace{1cm} & {\rm for} \ (\pi,\rho)+J/\Psi
   \rightarrow D+\bar{D} \label{mod}\\
& |M_f|^2 = 3 M_0^2 \hspace{1cm} & {\rm for} \ (\pi, \rho)+J/\Psi
   \rightarrow D^*+\bar{D}, D+\bar{D}^*, D^* + \bar{D}^* \nonumber\\
& |M_f|^2 = \frac{1}{3} M_0^2 \hspace{1cm} & {\rm for} \ (K,K^*)+J/\Psi
   \rightarrow D_s + \bar{D}, \bar{D}_s D \nonumber \\
& |M_f|^2 =  M_0^2 \hspace{1cm} & {\rm for} \ (K,K^*)+J/\Psi
   \rightarrow D_s + \bar{D}^*, \bar{D}_s D^*, D^*_s + \bar{D},
    \bar{D}^*_s D, \bar{D}^*_s D^* \nonumber
\label{mf}\end{eqnarray} involving a single parameter $M_0^2$ to
be fixed at SPS energies in comparison to the data of the NA50
Collaboration \cite{NA50b,NA50a}. The relative factors of 3 in
(\ref{mod}) are guided by the sum rule studies in \cite{korean}
which suggest that the cross section is increased whenever a
vector meson $D^*$ or $\bar{D}^*$ appears in the final channel
while another factor of 1/3 is introduced for each $s$ or
$\bar{s}$ quark involved. The factor $\left(
{(M_3+M_4)}/{\sqrt{s}} \right)^6 $ in (\ref{model}) accounts for
the suppression of binary channels with increasing $\sqrt{s}$ and
has been fitted to the experimental data for the reactions $\pi +
N \rightarrow \rho+N, \omega+N, \Phi+N, K^+ +\Lambda$ in Ref.
\cite{CaKo}. For simplicity we use the same matrix elements for
the dissociation of $\chi_c$ and $\Psi^\prime$ with mesons though
there is no fundamental reason why these matrix elements should be
the same. However, since we here concentrate only on the net
$J/\Psi$ absorption and production and not on the explicit
charmonium 'chemistry', this approximation should work out
reasonably well within the range of systematic uncertainties.

The advantage of the model introduced in (\ref{model}) is that
detailed balance for the binary reactions can be employed
strictly for each individual channel, i.e.
\begin{equation}
\label{balance}
 \sigma_{3+4 \rightarrow 1+2}(\sqrt{s}) =
 \sigma_{1+2 \rightarrow 3+4}(\sqrt{s})
\  \frac{(2S_1+1)(2S_2+1)}{(2S_3+1)(2S_4+1)} \
\frac{P_i^2}{P_f^2},
\end{equation}
 and the role of the backward reactions
($J/\Psi$+meson formation by $D+\bar{D}$ flavor exchange) can be
explored without introducing any additional parameter once $M_0^2$
is fixed. The uncertainty in the cross sections (\ref{model}) is
of the same order of magnitude as that in Lagrangian approaches
using e.g. $SU(4)_{flavor}$ symmetry \cite{Konew,Ko} since the
formfactors at the vertices are essentially unknown \cite{korean}.

As mentioned before, we fit the parameter $M_0^2$ to the $J/\Psi$
suppression data from the NA50 Collaboration for $Pb+Pb$
collisions at 160 A~GeV (cf. Section 4.1). For the value
$M_0^2$ = 0.13 fm/GeV$^2$ used below we end up with the $J/\Psi$
dissociation cross sections
\begin{equation}
\label{eqt}
\sigma_{J/\Psi + m \rightarrow X }(\sqrt{s}) =
\sum_c \sigma_{J/\Psi + m \rightarrow c}(\sqrt{s})
\end{equation}
displayed in Fig. \ref{bild6} with $\pi$, $\rho$, $K$ and $K^*$ mesons.
The summation over the final channel $c$ in (\ref{eqt}) includes all
binary channels compatible with charm quark and charge conservation.
Note, that for the comover absorption scenario essentially the regime
3.8 GeV $\leq \sqrt{s} \leq$ 4.8 GeV is of relevance (cf. Fig. 7.13 in
\cite{Cass99}) where the dissociation cross sections are on the level
of a few mb. We note, that the explicit channel $J/\Psi + \pi
\rightarrow D+ \bar{D}$, which has often been calculated in the
literature \cite{Konew,Ko,I2,I3} is below 0.7 mb in our model. A
somewhat more essential result is that the $J/\Psi$ dissociation cross
section with $\rho$-mesons is in the order of 5-7 mb as in the
calculations of Haglin \cite{Haglin} used before in Ref.
\cite{Cass01}, since this channel was found to dominate the $J/\Psi$
dissociation at SPS energies \cite{Cass99}. The explicit shape of the
cross sections is characterized by a rapid rise in $\sqrt{s}$ whenever
a new channel opens up. On the other hand, the channels with vector
mesons ($\rho, K^*$) are 'exothermal' and thus divergent at threshold.

The cross sections for the backward channels $D+\bar{D}, D+\bar{D}^*,
D^*+\bar{D}, D^*+\bar{D}^* \rightarrow J/\Psi$ + meson as well as the
channels involving $s$ or $\bar{s}$ quarks, i.e. $D_s+\bar{D},
D_s+\bar{D}^*, D^*_s+\bar{D}, D^*_s+\bar{D}^+ \rightarrow J/\Psi
+(K,K^*)$, then are fixed by detailed balance via (\ref{balance}). The
actual results for these channels -- summed up again over all possible
binary final states -- are displayed in Fig. \ref{bild7} separately for
the 'non-strangeness' (upper part) and 'strangeness' channels (lower
part) showing again divergent cross sections for 'exothermal' channels
like $D+\bar{D} \rightarrow J/\Psi + \pi$. Such divergent cross
sections arise in all 'exothermal' $S$-wave channels implying that
$D+\bar{D}$ or $D^*+\bar{D}$ mesons with low relative momentum have a
large cross section for $c$ and $\bar{c}$ quark exchange. In actual
transport calculations such divergent cross sections impose no problems
since the transition rates $\sim P_f \sigma_{3+4 \rightarrow 1+2}$
remain finite, as it is easily seen when inserting (\ref{model}) into
(\ref{balance}), since the divergent factor $P_f^2$ cancels out.
Furthermore, in the transport calculations an explicit cut in the total
cross sections of 120 mb is employed, which simulates the screening of
large cross sections at finite hadron density.

\subsection{Numerical implementation}

We recall that (as in Refs.  \cite{Cass01,Geiss99,CassKo,Cass97}) the
charm degrees of freedom are treated perturbatively and that initial
hard processes (such as $c\bar{c}$ or Drell-Yan production from $NN$
collisions) are 'precalculated' to achieve a scaling of the inclusive
cross section with the number of projectile and target nucleons as $A_P
\times A_T$ when integrating over impact parameter. To implement this
scaling we separate the production of the hard and soft processes: The
space-time production vertices of the $c\bar{c}$ pairs are
'precalculated' in each transport run by neglecting the soft processes,
i.e. the production of light quarks and assosiated mesons, and then
reinserted in the dynamical calculation at the proper space-time point
during the actual calculation that includes all soft processes. As
shown in Ref. \cite{Cass01} this prescription is very well in line with
Glauber calculations for the production of hard probes at fixed impact
parameter, too. We mention that this 'precalculation' of $c \bar{c}$
production might be modified at RHIC energies due to changes of the
gluon structure functions during the heavy-ion reaction or related
shadowing phenomena \cite{Strikman}. Such effects, however, are
expected to be of minor importance at RHIC energies (and below) and
will be discarded for our present study, that concentrates on the
balance between comover absorption and $J/\Psi$ reproduction channels.

Each open charm meson and charm vector meson is produced in the
transport calculation with a
weight $W_i$ given by the ratio of the actual production cross
section divided by the inelastic nucleon-nucleon cross section,
e.g.
\begin{equation}
W_i = \frac{\sigma_{NN \rightarrow J/\Psi +
x}(\sqrt{s})}{\sigma_{NN}^{inelas.}(\sqrt{s})}.
\end{equation}
In the transport simulation we follow the motion of the charmonium
pairs or produced $D, \bar{D}, D^*, \bar{D}^*$-mesons within the full
background of strings/hadrons by propagating them as free particles,
i.e. neglecting in-medium potentials, but compute their collisional
history with baryons and mesons or quarks and diquarks. For reactions
with diquarks we use the corresponding reaction cross section with
baryons multiplied by a factor of 2/3. For collisions with quarks
(antiquarks) we adopt half of the cross section for collisions with
mesons.

Furthermore, in addition to our previous studies
\cite{Cass00,Cass01,Cass97} the recreation of charmonia by channels
such as $D^*+ \bar{D} \rightarrow J/\Psi + \pi$ etc. is taken into
account in each individual run according to the cross sections
(\ref{balance}) with the weight of the produced charmonium states $k$
given by
\begin{equation}
W_k = W_i W_j,
\end{equation}
where $W_i, W_j$ are the individual weights of the open charm mesons.
The open charm mesons are not allowed to rescatter within a formation
time of 0.3 fm/c (in their rest frame) since a finite time is needed to
form their wave functions.  This formation time is not well known and
presently can only be estimated. Thus we checked -- by performing
calculations with formation times from 0.3 to 0.6 fm/c -- that the
physical statements (see below) remain robust. As commonly employed in
transport simulations, the open charm meson pairs, that stem from the
same interaction vertex, are not allowed to rescatter with each other
again unless an intermediate scattering has occurred.

\section{Nucleus-nucleus collisions}

\subsection{SPS energies}

We directly step on with the results for the charmonium suppression and
start with the system $Pb+Pb$ at 160 A~GeV to demonstrate that the
'late' comover dissociation model (\ref{model}) is approximately in
line with the data of the NA50 Collaboration. The corresponding
$J/\Psi$ suppression (in terms of the $\mu^+ \mu^-$ decay branch
relative to the Drell-Yan background from 2.9 -- 4.5 GeV invariant
mass) as a function of the transverse energy $E_T$ in $Pb~+~Pb$
collisions at 160 A~GeV is shown in Fig. \ref{bild8}. The solid line
(HSD'03) stands for the HSD result within the  comover absorption
scenario for the cross sections defined by (\ref{model}) while the
various data points reflect the different data releases from the NA50
Collaboration \cite{NA50aa,NA50bb,NA50b,NA50a}.  Note, that the 2002
data \cite{NA50_QM02} (lower part) no longer indicate the drop at the
highest $E_T$ (for analysis B) in line with the HSD calculations from
1997 \cite{Cass97} and the UrQMD results from 1999 \cite{Spieles2}
(dashed histogram). We mention that the present calculation (solid
line, HSD'03) agrees with the earlier calculations from Ref.
\cite{Cass97} (dotted line, HSD'97) very well except for the first
$E_T$-bin. Thus the cross sections presented in Fig. \ref{bild6} do not
lead to an overestimation of $J/\Psi$ suppression at SPS energies.
There might be alternative explanations for $J/\Psi$ suppression as
discussed in Refs.  \cite{Vogt99,Kojpsi,Rappnew,Geiss99,Satz99} and/or
further dissociation mechanism not considered here. However, for the
purposes of the present study it is sufficient to point out that the
cross sections displayed in Fig. \ref{bild6} most likely are upper
limits.

In order to provide some information on the relative production and
absorption channels for charmonia in these reactions we show the
calculated $J/\Psi$ rapidity distributions for 10\% central $Pb+Pb$
collisions at $\sqrt{s}$ = 17.3 GeV in Fig. \ref{bild9n}. The ordering
of the different lines is as follows: the upper dot-dot-dashed line
stands for the rapidity distribution of $J/\Psi$ mesons produced by
initial $BB$ collisions while the lowest dot-dashed line reflects the
rapidity distribution of $J/\Psi$ mesons from secondary $mB$ collisions
that are of minor importance at SPS energies. The dashed line
corresponds to the $J/\Psi$'s dissociated by baryons ($B$); this
absorption mechanism is denoted as 'conventional $J/\Psi$ attenuation'
by the NA50 Collaboration and also present in $p+A$ reactions. The
dotted line ('m abs.') gives the rapidity distribution for $J/\Psi$'s
dissociated with mesons ('comover absorption') while the full solid
line stands for the final $J/\Psi$ rapidity distribution.

As mentioned in Section 3, the model (\ref{model}) allows to calculate
the backward channels -- leading to $J/\Psi$ reformation by open charm
+ anticharm mesons -- without introducing any new parameter or
assumption. The result for the total $J/\Psi$ comover absorption rate
(solid histogram) in central $Pb~+~Pb$ collisions at 160 A~GeV is shown
in Fig. \ref{bild10n} in comparison to the $J/\Psi$ reformation rate
(dashed histogram) that includes all backward channels. Since the rates
differ by about 2 orders of magnitude, the backward rate for $J/\Psi$
formation can clearly be neglected at SPS energies even for central
$Pb+Pb$ reactions. This result is essentially due to the fact that the
expected multiplicity of open charm pairs is $\sim$ 0.12 in central
$Pb+Pb$ collisions at $\sqrt{s}$ = 17.3 GeV (according to the
calculations in Ref. \cite{Cass01}). Even in case of 'open charm
enhancement' (as suggested in Ref. \cite{NA50d}) by a factor $\sim$ 3,
where the $J/\Psi$ reformation rate would increase by a factor $\sim$
9, the backward channels still could be neglected.

Since the 'comover' dissociation cross sections employed should be
regarded as upper limits, we conclude that no chemical equilibration
between mesons, open charm mesons and charmonia is achieved dynamically
at SPS energies. Note, however, that the transverse mass $M_T$ spectra
for all mesons including open charm and charmonia from central $Pb+Pb$
collisions scale according to the HSD calculations (cf. Fig. 18 of Ref.
\cite{Cass01}), if final state elastic scatterings are omitted. Thus
statistical model fits still should work for the different hadron
abundancies.

\subsection{RHIC energies}

For central $Au+Au$ collisions at $\sqrt{s}$ = 200 GeV, however,
the multiplicity of open charm pairs should be $\sim$ 16, i.e. by
about 2 orders of magnitude larger, such that a much higher
$J/\Psi$ reformation rate ($\sim N_{c\bar{c}}^2$) is expected at
RHIC energies (cf. Ref. \cite{Rappnew}). In Fig. \ref{bild11n} we
display the total $J/\Psi$ comover absorption rate (solid
histogram)  in comparison to the $J/\Psi$ reformation rate (dashed
histogram) as a function of time in the center-of-mass frame.
Contrary to Fig. \ref{bild10n} now the two rates become comparable
for $t \geq$ 4-5 fm/c and suggest that at the full RHIC energy of
$\sqrt{s}$ = 200 GeV the $J/\Psi$ comover dissociation is no
longer important since the charmonia dissociated in this channel
are approximately recreated in the backward channels. Accordingly,
the $J/\Psi$ dissociation at RHIC should be less pronounced  than
at SPS energies. Moreover, there is even a small excess of
$J/\Psi$ formation by $D+\bar{D}$ reactions in the first 2 fm/c
qualitatively in line with AMPT calculations by Zhang et al.
\cite{zhang02}.

In order to provide some information on the relative production and
absorption channels for charmonia in these reactions we show -- in
analogy to Fig. \ref{bild9n} -- the calculated $J/\Psi$ rapidity
distributions for 12\% central $Au+Au$ collisions at $\sqrt{s}$ = 200
GeV in the upper part of Fig. \ref{bild12n}. The ordering of the
different lines is as follows: the upper dot-dot-dashed line stands for
the rapidity distribution of $J/\Psi$ mesons produced by initial $BB$
collisions while the lowest dot-dashed line reflects the rapidity
distribution of $J/\Psi$ mesons from secondary $mB$ collisions that are
of minor importance also at RHIC energies. The dashed line corresponds
to the $J/\Psi$'s dissociated by baryons ($B$) and corresponds to the
 'conventional $J/\Psi$ attenuation'. This distribution is
approximately the same as the recreation of $J/\Psi$'s from $D+\bar{D}$
annihilation (thin solid line with open circles).  The dotted line ('m
abs.') gives the rapidity distribution for $J/\Psi$'s dissociated with
mesons ('comover absorption'); it is slightly lower than the
$D+\bar{D}$ recreation channel.  The full solid line stands for the
final $J/\Psi$ rapidity distribution which is about a factor of $\sim
3$ lower than the primary production from $BB$ collisions. Since all
distributions (within statistics) are practically flat for $|y_{cm}|
\leq 2$ no strong sensitivity of the $J/\Psi$ survival probability is
expected for different rapidity cuts in this interval around
midrapidity.

We additionally comment on results of HSD calculations that have been
performed under the assumption of initial $J/\Psi$ 'melting' by color
screening in a QGP phase as advocated in Refs.  \cite{Satz,Satznew}. To
this aim we have 'deleted' all charmonia created initially from primary
$BB$ collisions in the calculation, but evolved the system in time with
the same production and absorption cross sections as before. The
resulting final $J/\Psi$ rapidity distribution for central $Au+Au$
collisions at $\sqrt{s}$ = 200 GeV is shown in the lower part of Fig.
\ref{bild12n} by the dashed line in comparison to the final $J/\Psi$
rapidity distribution from the upper part of the figure (solid line).
The comparison demonstrates that even in case of complete initial
charmonium dissociation a finite amount of $J/\Psi$'s should be seen
experimentally, which is roughly half of the yield expected from the
full calculations and essentially due to the $D+\bar{D}$ production
channels. Since the latter cross sections are upper estimates, the
$J/\Psi$ yield (dashed line in Fig. \ref{bild12n}) also has to be
considered as an upper limit in this case.

A note of caution should be added in context with Fig.
\ref{bild12n} since the actual rapidity distributions might change
quantitatively when including a more refined model for the matrix
elements in (7) especially for the $\chi_c$ and $\Psi^\prime$
states. Furthermore, in-medium modifications (or self-energy
corrections) of the open charm mesons (and charmonia) should
change the final rapidity distributions to some extent since a
lowering of $D,\bar{D}$ masses leads to an increase of $J/\Psi$ +
meson absorption rates and a decrease of the backward channel
rates \cite{zhang02}. For constant matrix elements as in (7) these
modifications directly result from an enhanced phase space for
absorption and a reduced invariant energy for the backward
channels. On the other hand, for enhanced $D,\bar{D}$ masses in
the medium the $J/\Psi$ + meson absorption rates will be lowered
and the backward channels be enhanced accordingly. As argued in
Ref. \cite{golub02} charmonium spectroscopy in $\bar{p}$ induced
reactions on nuclei might shed some further light on this
presently open issue. Nevertheless, our actual results for the
$J/\Psi$ reformation by open charm + anticharm mesons are in
qualitative and even quantitative agreement with the independent
transport studies in Ref. \cite{zhang02} that also demonstrate a
net reduction of $J/\Psi$ mesons relative to the extrapolations
from $pp$ collisions with the number of binary collisions.

We now turn back again to the HSD results for the full
calculations. To quantify the final $J/\Psi$ suppression in
$Au+Au$ collisions at RHIC we show in Fig. \ref{bild13n} the
calculated $J/\Psi$ survival probability $S_{J/\Psi}$ defined as
\begin{equation} \label{supp} S_{J/\Psi} =
\frac{N^{J/\Psi}_{fin}}{N^{J/\Psi}_{BB}}, \end{equation} where $
N^{J/\Psi}_{fin}$ and $N^{J/\Psi}_{BB}$ denote the final number of
$J/\Psi$ mesons and the number of $J/\Psi$'s produced initially by
$BB$ reactions, respectively.  In Fig. \ref{bild13n} the quantity
(\ref{supp}) is displayed as a function of the transverse energy
$E_T$ -- in units of the transverse energy at impact parameter
$b=1$ fm -- for $Au+Au$ collisions with (solid line) and without
inclusion of the backward channels (dash-dotted line). In fact,
the dash-dotted line is (within statistics) identical to the
previous calculation in Ref. \cite{Cass00} demonstrating a
considerable $J/\Psi$ 'comover' suppression for central
collisions. When including the reformation channels this
suppression is substantially reduced and leads to a less effective
dissociation of charmonia than at SPS energies (middle dashed
line). Furthermore, we observe that $S_{J/\Psi} \leq 1$ for all
centralities and thus no $J/\Psi$ enhancement relative to the
primary $BB$ production is found from our calculations as claimed
in the statistical models of Refs. \cite{Rafelski,Go1,Peter}. We
also like to recall that the charmonium 'melting scenario'
advocated in Ref. \cite{Satz99} should lead to a step-like $E_T$
dependence of $S_{J/\Psi}$ due to a successive melting of the
$\chi_c$ and $J/\Psi$ and an almost complete disappearance of
$J/\Psi$'s for central collisions. Moreover, as shown in Refs.
\cite{Gorenst1,Gorenst2}, statistical models on the partonic or
even hadronic level lead to very different predictions for the
$J/\Psi$ multiplicity as a function of centrality in $Au+Au$
collisions at $\sqrt{s}$ = 200 GeV.  Since at RHIC energies the
predictions of the 'comover' approach, the statistical models and
the 'melting scenario' are substantially different, experiment
should clearly decide about the adequacy of the concepts involved.

The preliminary data of the PHENIX Collaboration \cite{PHENIX}
allow for a first glance at the situation encountered in $Au+Au$
collisions at $\sqrt{s}$ = 200 GeV. In order to compare with the
preliminary data we have performed a rapidity cut $|y_{cm}| \leq
2$ in the calculations. In Fig. \ref{bild14n} the $J/\Psi$
multiplicity per binary collision (times the branching ratio $B$)
is shown as a function of the number of participating nucleons
$A_{part}$ in comparison to the data at midrapidity.  Whereas our
transport results give a monotonous decrease of the $J/\Psi$ yield
(per binary collision) with centrality, the statistical charm
coalescence model of Gorenstein et al. \cite{Gorenst2} predicts an
increase by about 20\% from $A_{part}$ = 100 to 380.  Since the
statistics (and binning in $A_{part}$) is quite limited so far on
the experimental side, no final conclusion can presently be drawn,
however, the data neither suggest a dramatic enhancement of
$J/\Psi$ production nor a complete 'melting' of the charmonia in
the QGP phase.

\section{Summary}

In this work we have performed a first comparison of results from HSD
transport calculations on meson, baryon, antibaryon production and
elliptic flow with the (preliminary) data for $ Au~+~Au$ collisions at
$\sqrt{s}$ = 200 GeV from the PHOBOS, BRAHMS and PHENIX Collaborations.
The HSD transport approach, which is based on quark, diquark, string
and hadronic degrees of freedom, is found to  give a quite reasonable
description of the different observables studied in this work. Only the
elliptic flow $v_2$ is underestimated closer to midrapidity --
quantitatively in line with the hadron-string cascade calculations in
Ref. \cite{Sahu02} -- indicating that there might be 'extra pressure'
being generated in the 'prehadronic phase'.

On the other hand, hard probes such as charmonia and open $D$-meson
pairs are expected to be sensitive to the initial phase of high energy
density where charmonia might be 'melting' according to the scenario
advocated in Ref. \cite{Satz99}, their formation be suppressed due to
plasma screening \cite{CMKO} or absorbed early by neighboring strings
\cite{Geiss99}. However, charmonia might also be generated in a
statistical fashion at the phase boundary between the QGP and an
interacting hadron gas such that their abundance could be in
statistical (chemical) equilibrium with the light and strange hadrons
\cite{Marek1,Peter}. The latter picture  is expected to lead not to a
suppression but to an enhancement of $J/\Psi$ mesons at the full RHIC
energy if compared to the scaled $J/\Psi$ multiplicity from $pp$
collisions \cite{Rafelski}. We recall that the 'hadronic comover'
dissociation concept has lead to a $\sim$ 90 \% $J/\Psi$ suppression in
central $Au + Au$ collisions at $\sqrt{s}$ \cite{Cass00} due to the
high meson densities encountered, however, as pointed out in
\cite{Cass00}, the latter calculations had been performed without
including the backward $D + \bar{D} \rightarrow J/\Psi$ + meson
channels thus violating 'detailed balance'.

The focus of this work has been to show the dynamical effects from
the backward channels for charmonium reproduction by $D + \bar{D}$
channels employing detailed balance on a microscopic level. To
this aim we have formulated a simple phase-space model for the
individual charmonium dissociation channels with a single free
parameter $M_0^2$ (cf. Section 3), which we have fixed at SPS
energies in comparison to the $J/\Psi$ suppression data of the
NA50 Collaboration.  In fact, the results for the charmonium
suppression are practically the same as in the previous HSD
transport calculations \cite{Cass00,Cass01,Cass97}. From our
dynamical calculations we find that the charmonium recreation by
the backward channels plays no role at SPS energies (cf. Fig. 10),
however, becomes substantial in $Au + Au$ collisions at $\sqrt{s}$
= 200 GeV and even is slightly larger than the 'comover'
absorption channel. This leads to the final result that the total
$J/\Psi$ suppression as a function of centrality is less
pronounced than at SPS energies, where the backward channels play
no role. Furthermore, even in case that all directly produced
$J/\Psi$ mesons are not formed as a mesonic state (e.g. due to
color screening), a sizeable amount of charmonia is found
asymptotically due to the $D+\bar{D} \rightarrow J/\Psi$ + meson
channels which is almost quantitatively in line with the AMPT
calculations in Ref. \cite{zhang02} for central $Au+Au$ collisions
at $\sqrt{s}$ = 200 GeV. Since the cross sections for $J/\Psi$ +
meson absorption employed in this work have to be considered as
upper limits, the charmonium reformation by $D+\bar{D} \rightarrow
J/\Psi$ + meson channels should be lower than the $J/\Psi$ cross
section expected from binary scaling of $pp$ reactions. The
preliminary data of the PHENIX Collaboration \cite{PHENIX} are
compatible with our full transport calculations (cf. Fig.
\ref{bild14n}), however, improved statistics and also data for
light systems such as $Ne+Ne$ and $Ag+Ag$ will be necessary to
clarify the issue of charmonium suppression experimentally.

\section*{Acknowledgements}
\begin{figure}[h]
\centerline{\psfig{file=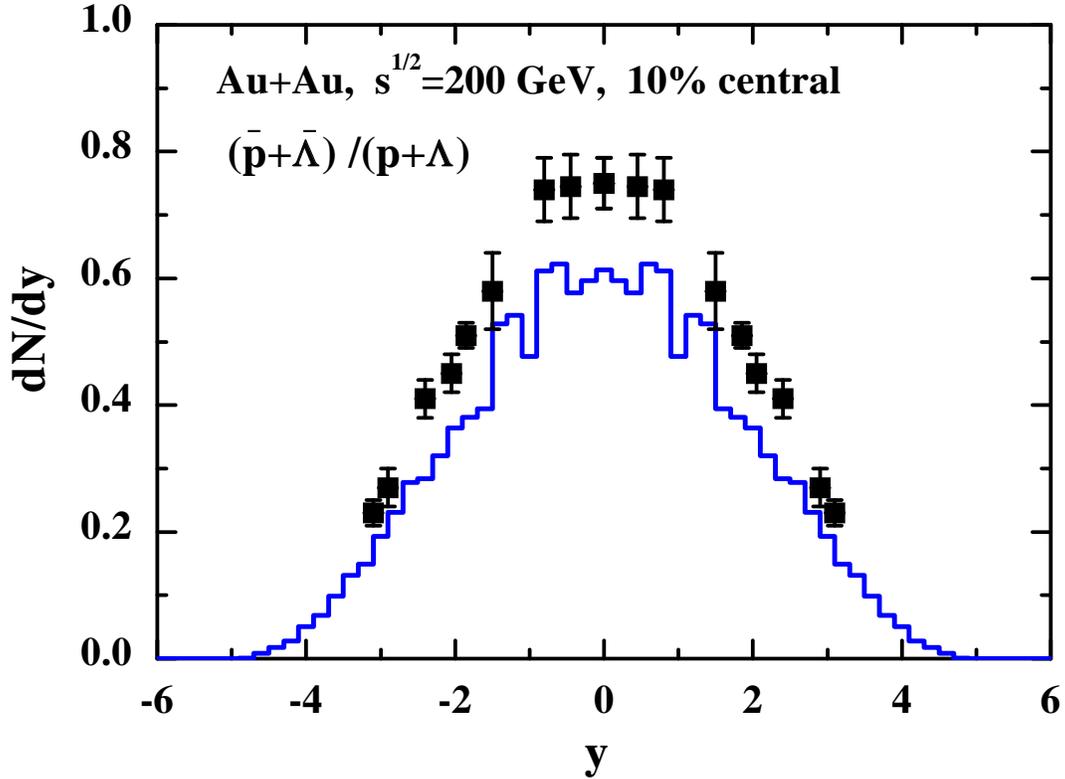,width=14cm}}
\vspace*{5mm}
\caption{The $(\bar{p} + \bar{\Lambda})/(p + \Lambda)$ ratio in 10\%
central $Au+Au$ collisions at $\sqrt{s}$ = 200 GeV as a function of
rapidity $y$ in comparison to the $\bar{p}/p$ data from the BRAHMS
Collaboration \protect\cite{BRAHMS2}. Note, that the experimental data
include some unknown fraction of $\Lambda$ and $\bar{\Lambda}$ decays
such that the comparison suffers from a 5-10\% systematic uncertainty.}
\label{bild2} \end{figure}

\begin{figure}[h]
\centerline{\psfig{file=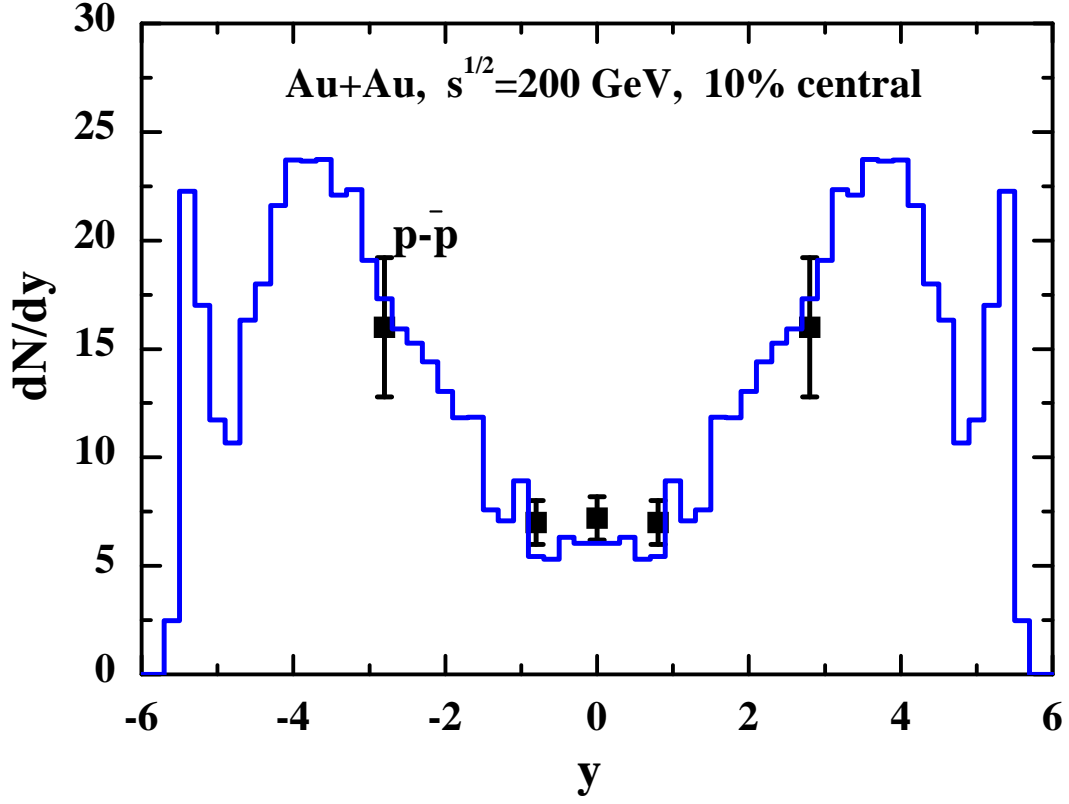,width=14cm}}
\vspace*{5mm}
\caption{The  net proton $(p-\bar{p})$ rapidity distribution in central
$Au+Au$ collisions at $\sqrt{s}$ = 200 GeV in comparison to the
preliminary data of the BRAHMS Collaboration \protect\cite{BRAHMS3} for
the same event class as in Fig. \protect\ref{bild2}. Note, that the experimental data
include some unknown fraction of $\Lambda$ and $\bar{\Lambda}$
decays such that the 'real' $(p-\bar{p})$ rapidity distribution should be slightly lower.}
\label{bild3} \end{figure}

\begin{figure}[h]
\centerline{\psfig{file=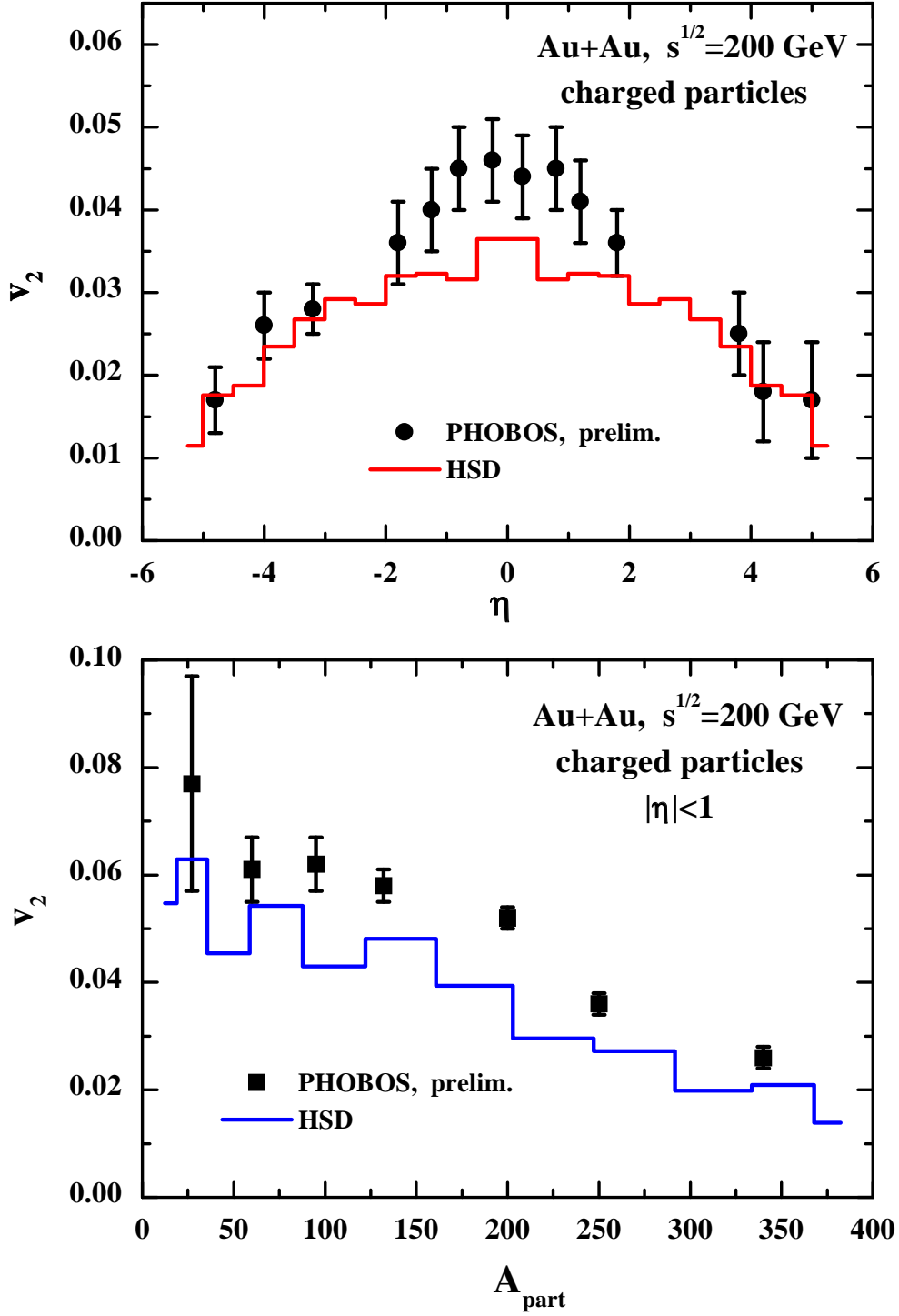,width=13cm}}
\vspace*{5mm}
\caption{The  calculated elliptic flow $v_2$ for charged hadrons (solid
lines) as a function of pseudorapidity $\eta$ (upper part) and as a
function of the number of 'participating nucleons' $A_{part}$ for
$|\eta| \leq$ 1 (lower part) for $Au+Au$ collisions at $\sqrt{s}$ = 200
GeV in comparison to the preliminary 'hit-based analysis' data of the
PHOBOS Collaboration \protect\cite{PHOBOS1}.}
\label{bild4} \end{figure}

\begin{figure}[h]
\centerline{\psfig{file=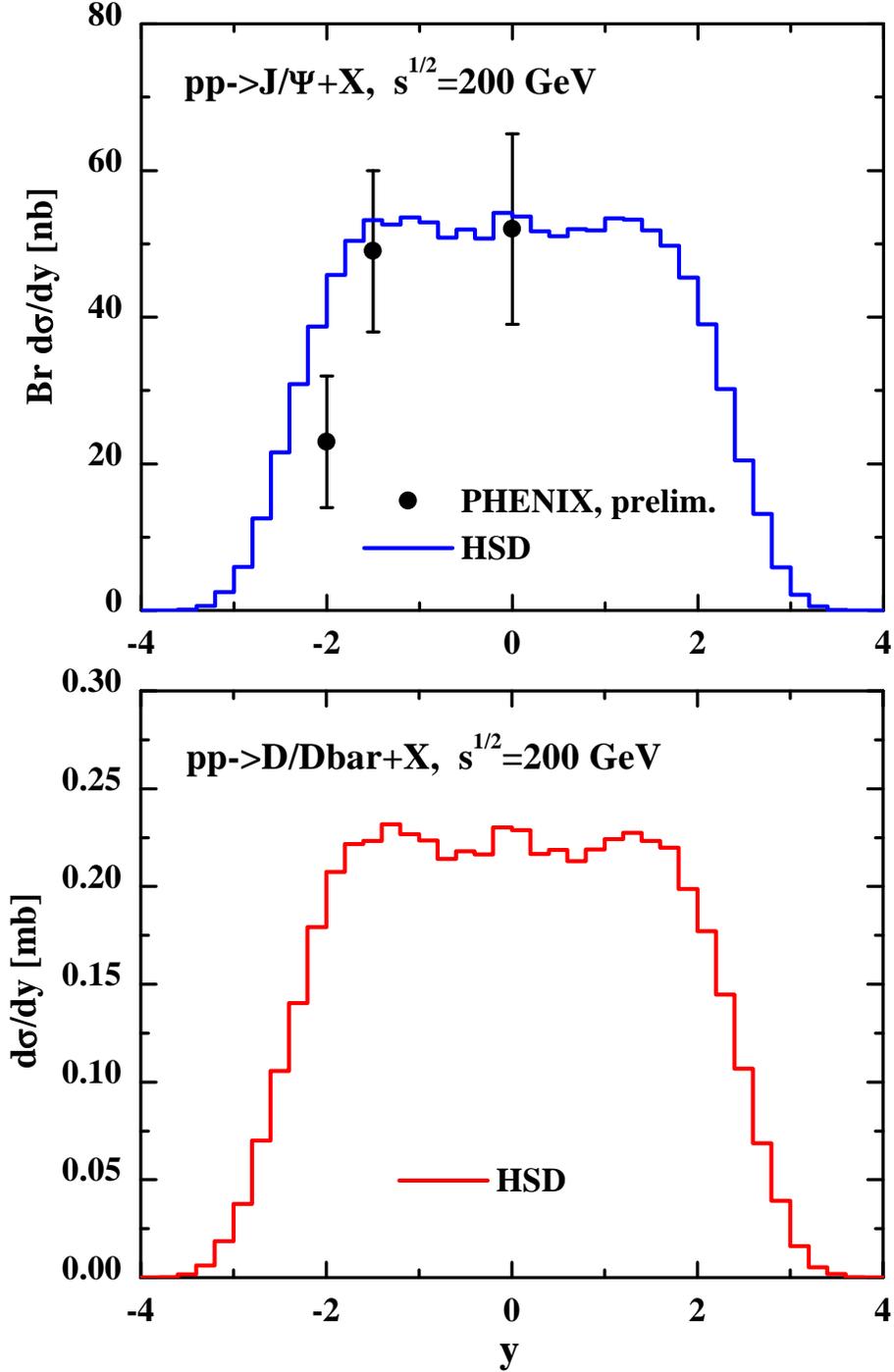,width=12cm}}
\vspace*{5mm}
\caption{The  calculated rapidity distribution for $J/\Psi$ mesons
(upper part, multiplied by the branching to dileptons) and all open
charm mesons (lower part) from $pp$ collisions at $\sqrt{s}$ = 200 GeV
in comparison to the preliminary data from the PHENIX Collaboration
\protect\cite{PHENIX} for $J/\Psi + X$.  The $D+\bar{D}$ pair rapidity
distribution is obtained by dividing the result in the lower part by a
factor of $\sim 2$.}
\label{bild5} \end{figure}

\begin{figure}[h]
\centerline{\psfig{file=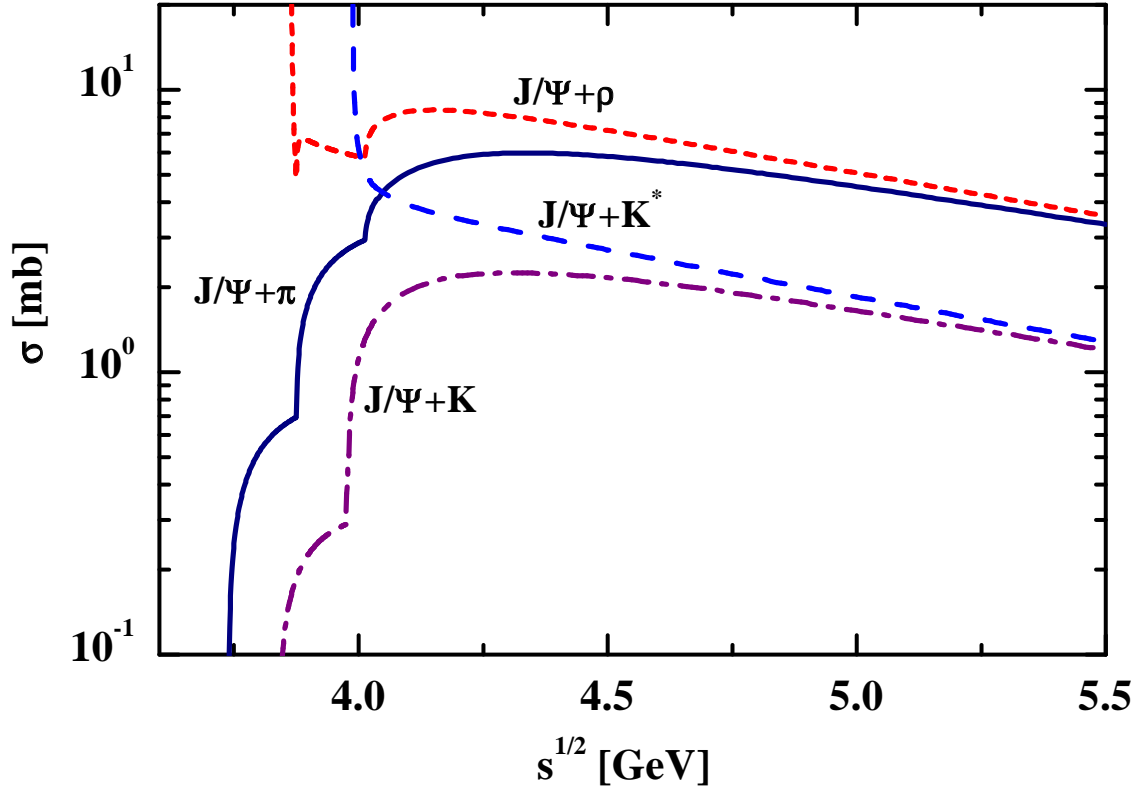,width=15cm}}
\vspace*{5mm}
\caption{The $J/\Psi$ dissociation cross sections  with $\pi$, $\rho,
K$ and $K^*$ mesons as specified in Section 3.}
\label{bild6} \end{figure}

\begin{figure}[h]
\centerline{\psfig{file=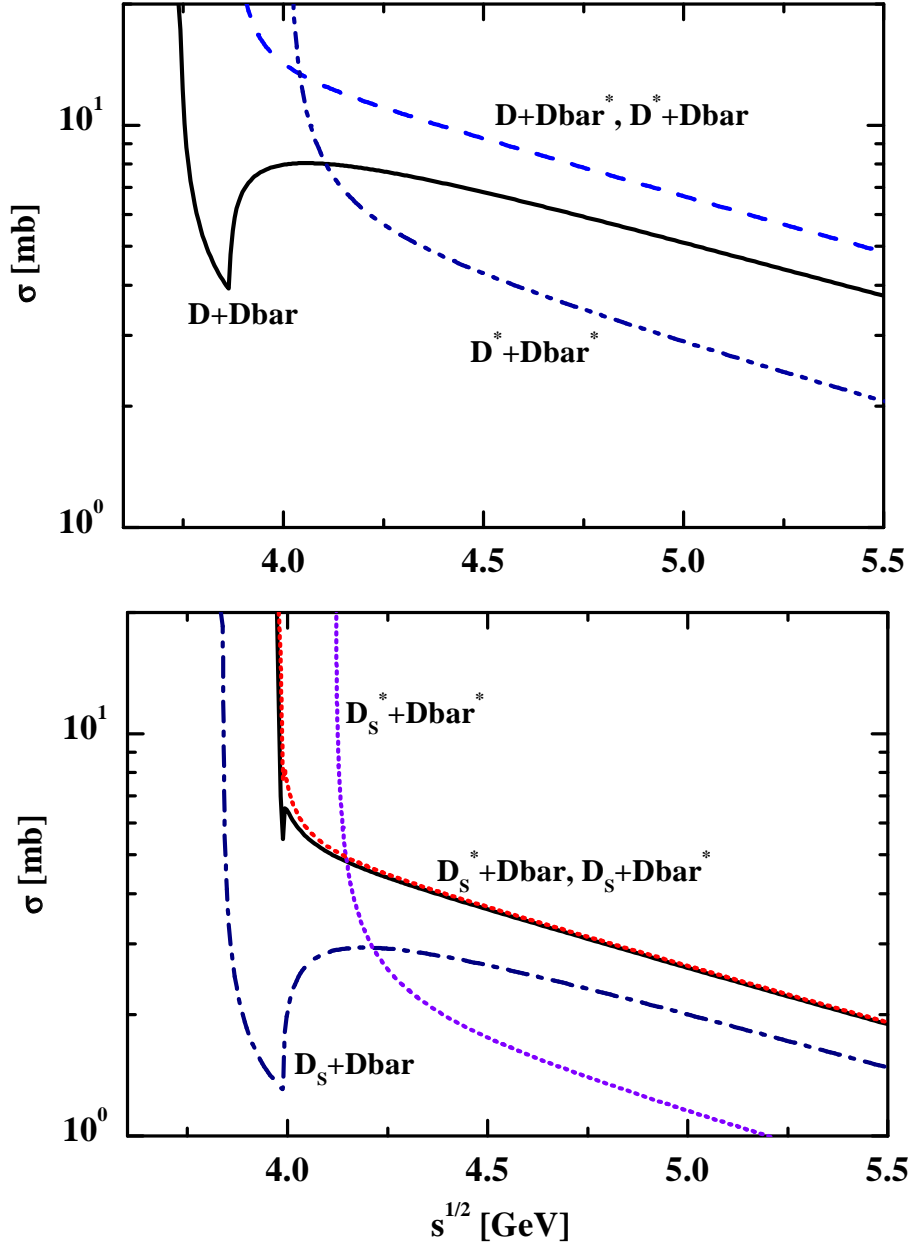,width=12cm}}
\vspace*{5mm}
\caption{The cross sections for the channels $D+\bar{D}, D+\bar{D}^*,
D^*+\bar{D}, D^*+\bar{D}^* \rightarrow J/\Psi$ + meson (upper part) and
the channels involving $s$ or $\bar{s}$ quarks $D_s+\bar{D}$,
$D_s+\bar{D}^*$, $D^*_s+\bar{D}$, $D^*_s+\bar{D}^+ \rightarrow J/\Psi
+(K,K^*)$ (lower part) as a function of the invariant energy $\sqrt{s}$
according to the model described in Section 3.}
\label{bild7} \end{figure}

\begin{figure}[t]
\centerline{\psfig{figure=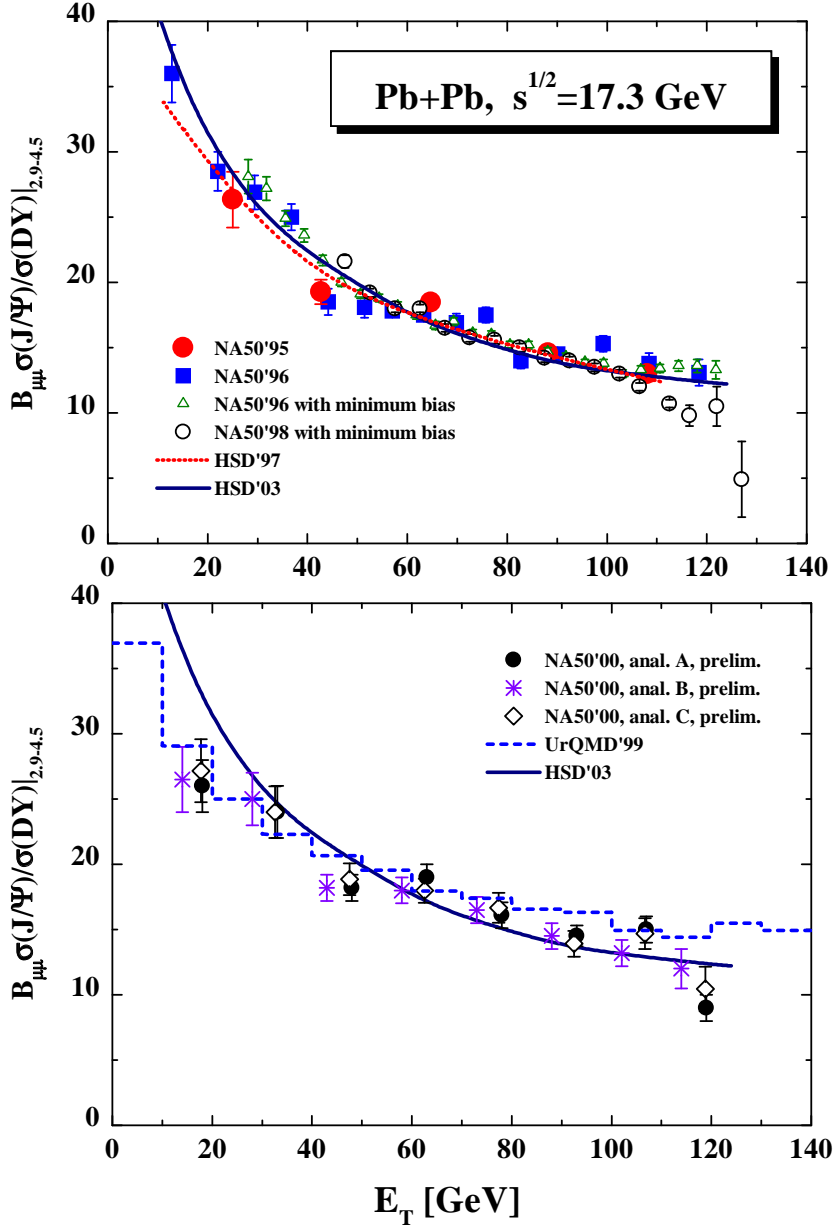,width=11cm}} \vspace*{5mm}
\caption{The $J/\Psi$ suppression (in terms of the $\mu^+ \mu^-$
decay branch relative to the Drell-Yan background from 2.9 -- 4.5
GeV invariant mass) as a function of the transverse energy $E_T$
in $Pb~+~Pb$ collisions at 160 A~GeV. The solid line (HSD'03)
stands for the HSD result within the 'late' comover absorption
scenario presented in Section 3 while the dotted line (HSD'97)
reflects the earlier calculation from Ref. \protect\cite{Cass97}.
Upper part: the full dots stand for the NA50 data from 1995, the
full squares for the 1996 data, the open triangles for the 1996
data with minimum bias while the open circles represent the 1998
data adopted from Refs. \protect\cite{NA50aa,NA50bb,NA50b,NA50a}.
Lower part: the open and full symbols indicate the preliminary
NA50 data from 2000 (analysis A, B and C)
\protect\cite{NA50_QM02}. The dashed histogram  is the UrQMD
result from Ref. \protect\cite{Spieles2}.} \label{bild8}
\end{figure}

\begin{figure}[h]
\centerline{\psfig{file=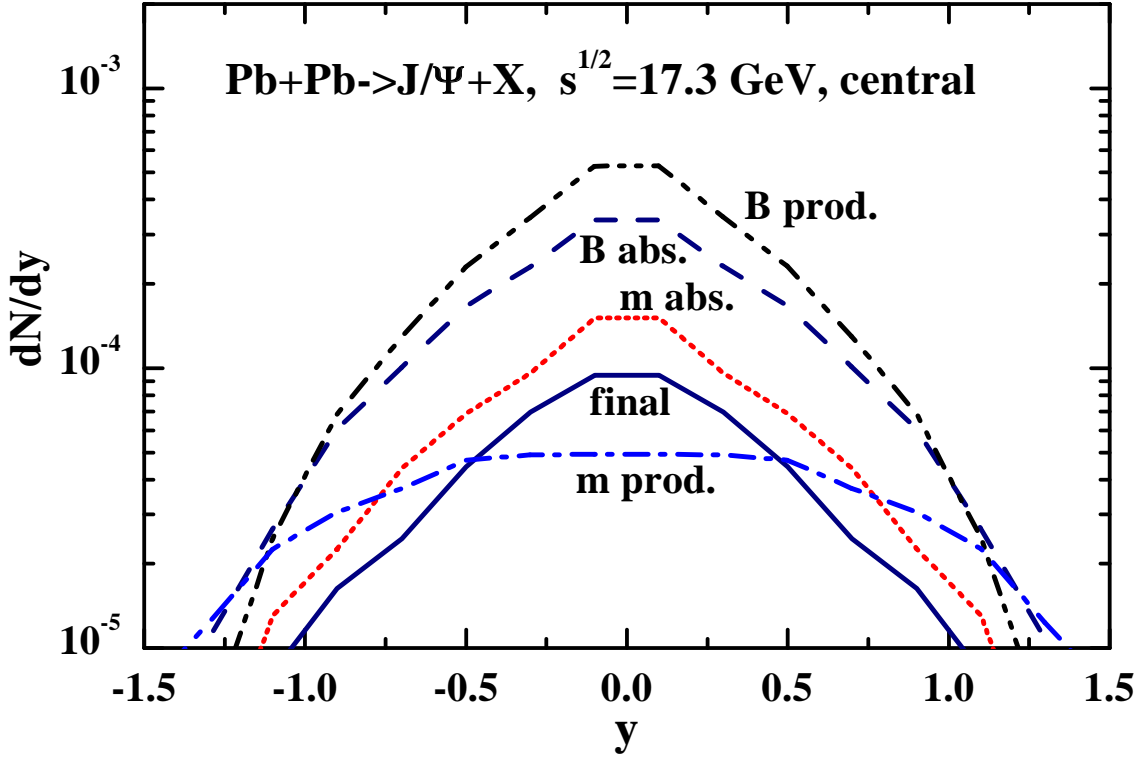,width=15cm}}
\vspace*{5mm}
\caption{Calculated $J/\Psi$ rapidity distributions for central $Pb+Pb$
collisions at $\sqrt{s}$ = 17.3 GeV. The ordering of the different
lines is as follows: the upper dot-dot-dashed line stands for the
rapidity distribution of $J/\Psi$ mesons produced by initial $BB$
collisions while the lowest dot-dashed line reflects the rapidity
distribution of $J/\Psi$ mesons from $mB$ collisions. The dashed line
corresponds to the $J/\Psi$'s dissociated by baryons ($B$) and the
dotted line shows the $J/\Psi$'s dissociated by mesons ($m$). The full
solid line gives the final $J/\Psi$ rapidity distribution.}
 \label{bild9n} \end{figure}

\begin{figure}[h]
\centerline{\psfig{file=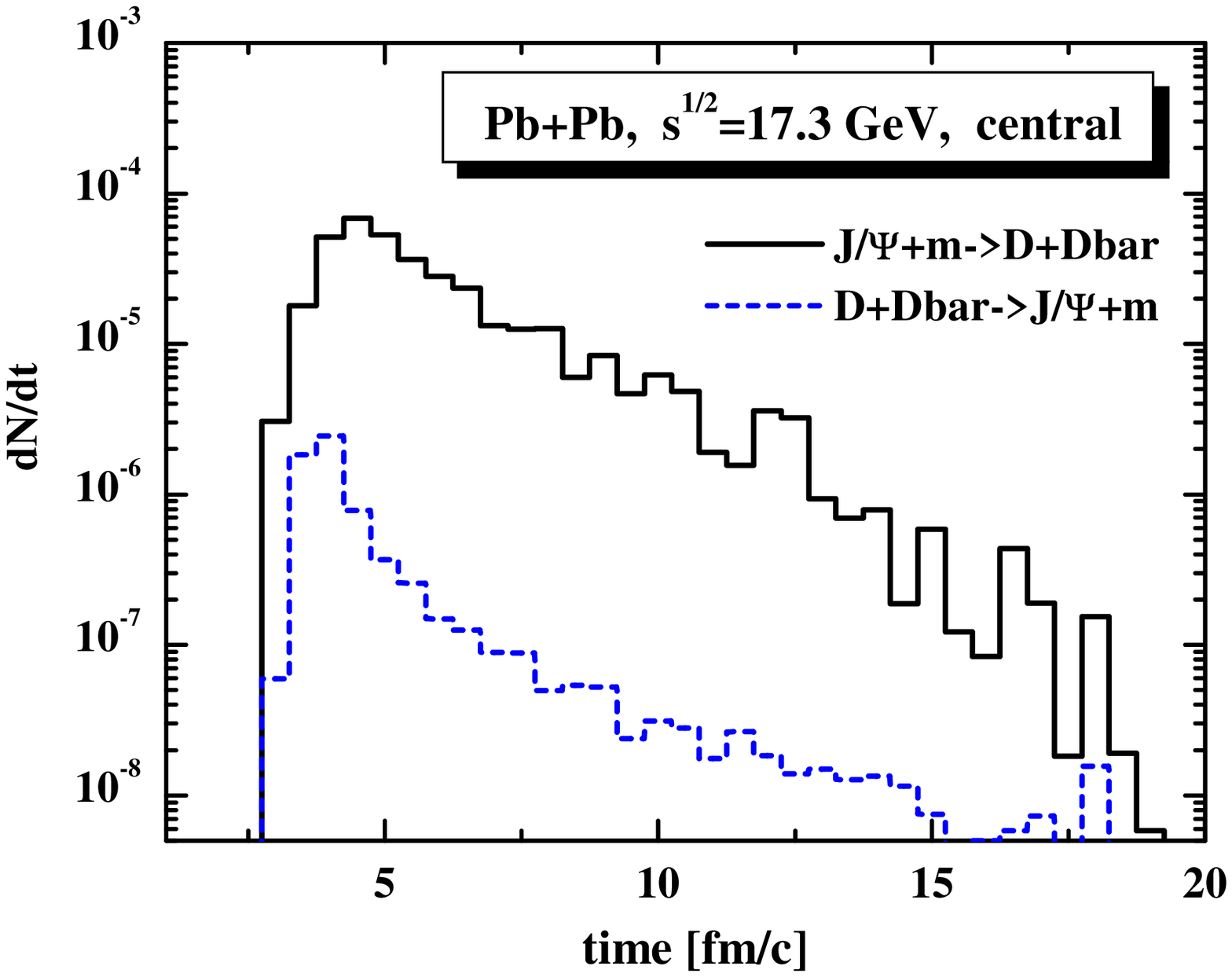,width=15cm}}
\vspace*{5mm}
\caption{The  calculated rate of $J/\Psi$ dissociation reactions with
mesons (solid histogram) for central $Pb+Pb$ collisions at $\sqrt{s}$ =
17.3 GeV in comparison the rate of backward reactions of open charm
pairs to $J/\Psi$ + meson (dashed histogram) according to the model
specified in Section 3.}
\label{bild10n} \end{figure}

\begin{figure}[h]
\centerline{\psfig{file=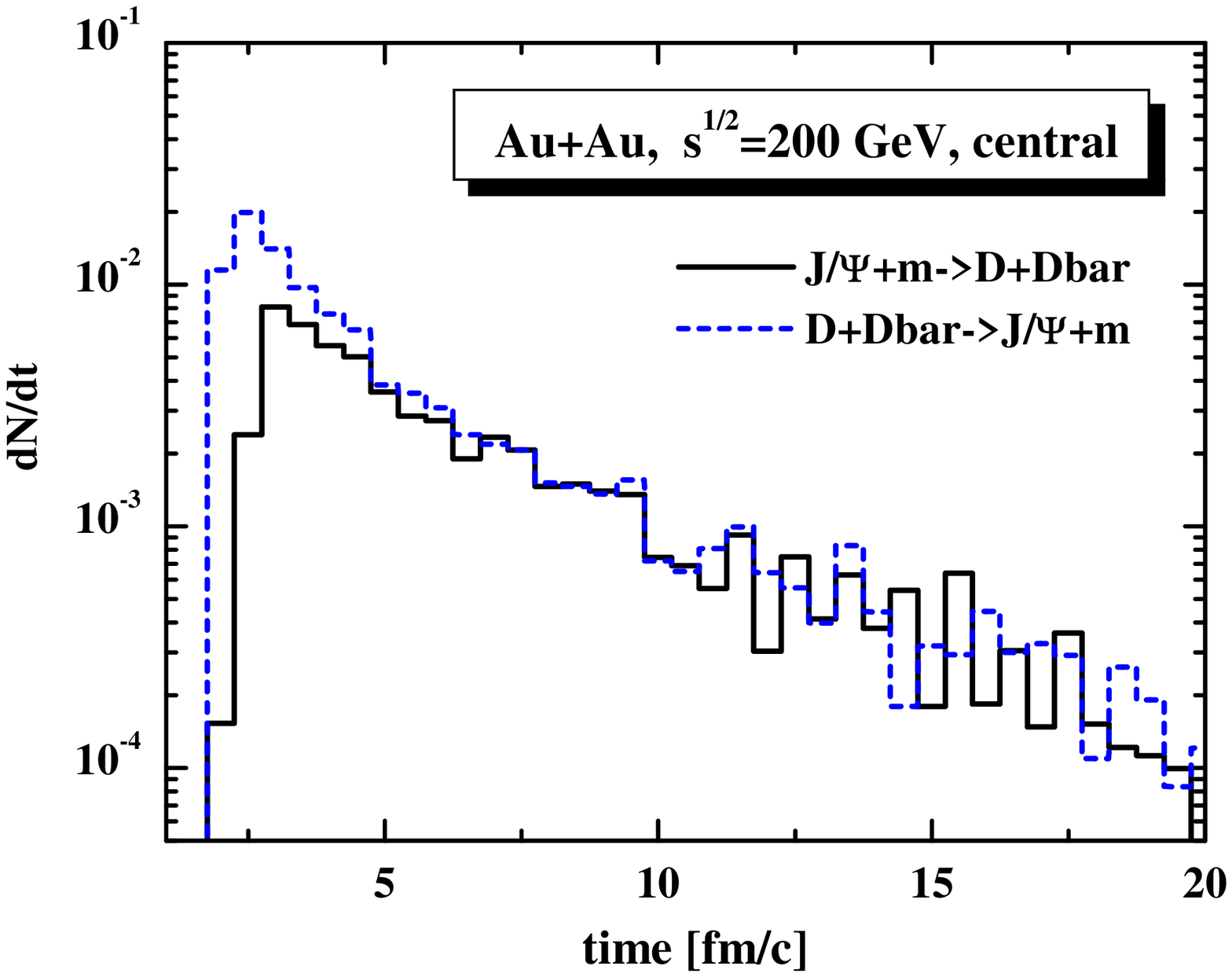,width=15cm}}
\vspace*{5mm}
\caption{The  calculated rate of $J/\Psi$ dissociation reactions with
mesons (solid histogram) for central $Au+Au$ collisions at $\sqrt{s}$ =
200 GeV in comparison the rate of backward reactions of open charm
pairs to $J/\Psi$ + meson (dashed histogram) according to the model
specified in Section 3.}
\label{bild11n} \end{figure}

\begin{figure}[h]
\centerline{\psfig{file=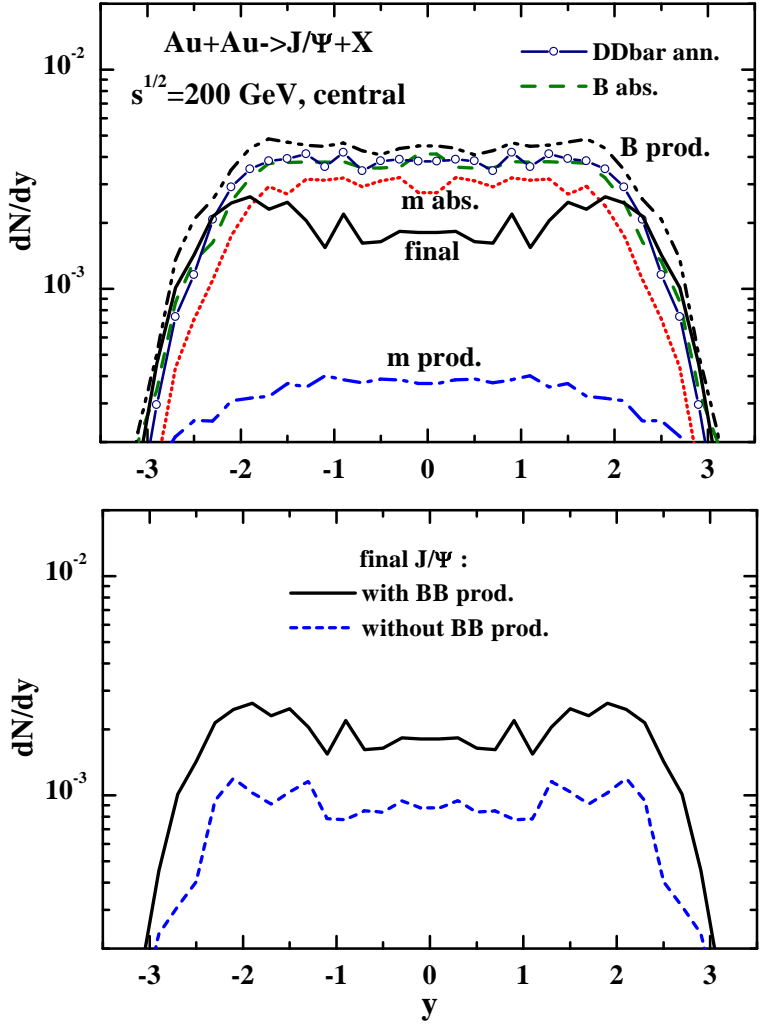,width=10cm}} \vspace*{5mm}
\caption{Calculated $J/\Psi$ rapidity distributions for 10\%
central $Au+Au$ collisions at $\sqrt{s}$ = 200 GeV. The ordering
of the different lines in the upper part is as follows: the upper
dot-dot-dashed line stands for the rapidity distribution of
$J/\Psi$ mesons produced by initial $BB$ collisions while the
lowest dot-dashed line reflects the rapidity distribution of
$J/\Psi$ mesons from $mB$ collisions. The dashed line corresponds
to the $J/\Psi$'s dissociated by baryons ($B$); this distribution
is approximately the same as the recreation of $J/\Psi$'s from
$D+\bar{D}$ annihilation (thin solid line with open circles). The
dotted line ('m abs.') shows the $J/\Psi$'s dissociated by mesons
($m$), which is slightly lower than the $D+\bar{D} \rightarrow
J/\Psi + $meson recreation channel. The full solid line gives the
final $J/\Psi$ rapidity distribution. Lower part: The solid line
is identical to the final $J/\Psi$ rapidity distribution from the
upper part whereas the dashed line is obtained from HSD
calculations assuming that all charmonia produced from initial
$BB$ collisions are  'melted' in a possible QGP phase (see text).}
\label{bild12n}
\end{figure}

\begin{figure}[h]
\centerline{\psfig{file=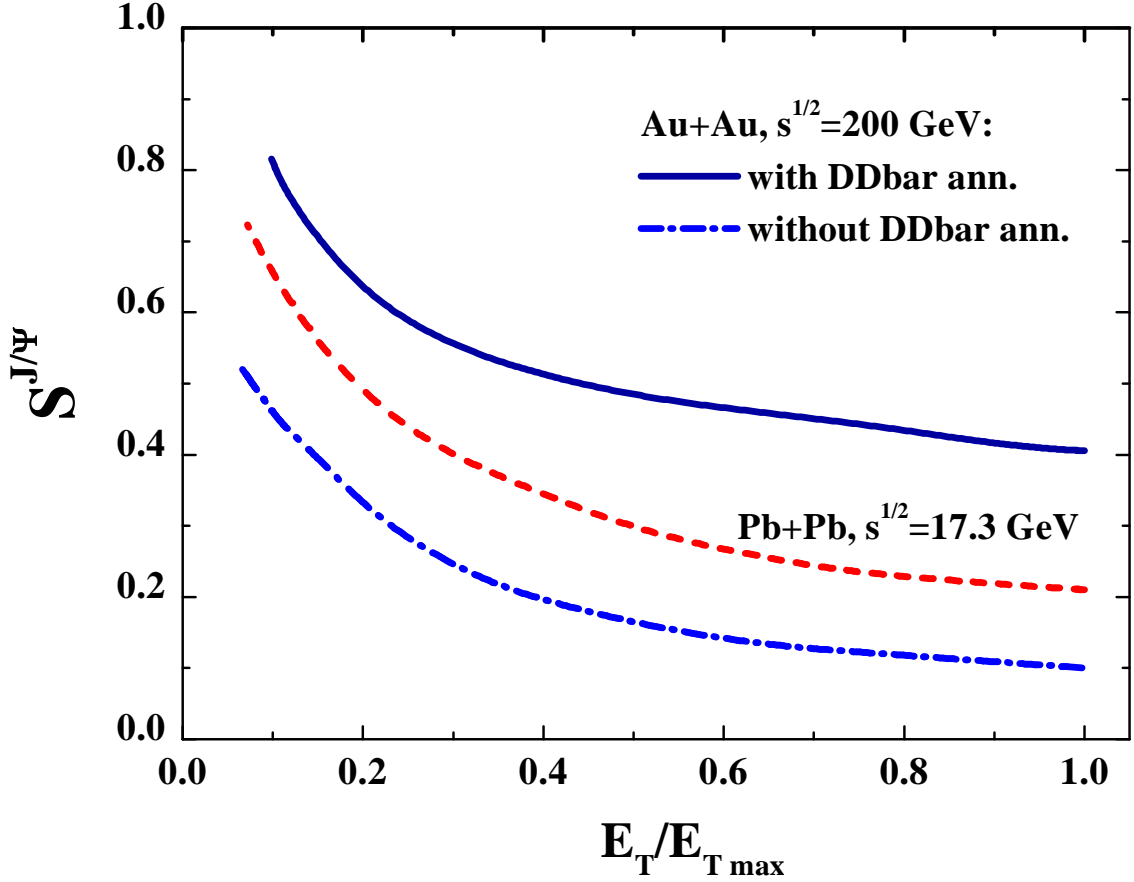,width=15cm}} \vspace*{5mm}
\caption{The calculated $J/\Psi$ survival probability $S_{J/\Psi}$
as a function of the transverse energy - in units of the
transverse energy at impact parameter $b=1$ fm -- for $Au+Au$
collisions with (solid line) and without inclusion of the backward
channels (lower dot-dashed line). The dashed line (middle) shows
the result from Fig. 8 for the same quantity in $Pb+Pb$ collisions
at $\sqrt{s}$ = 17.3 GeV for
 comparison.}
 \label{bild13n} \end{figure}

\begin{figure}[h]
\centerline{\psfig{file=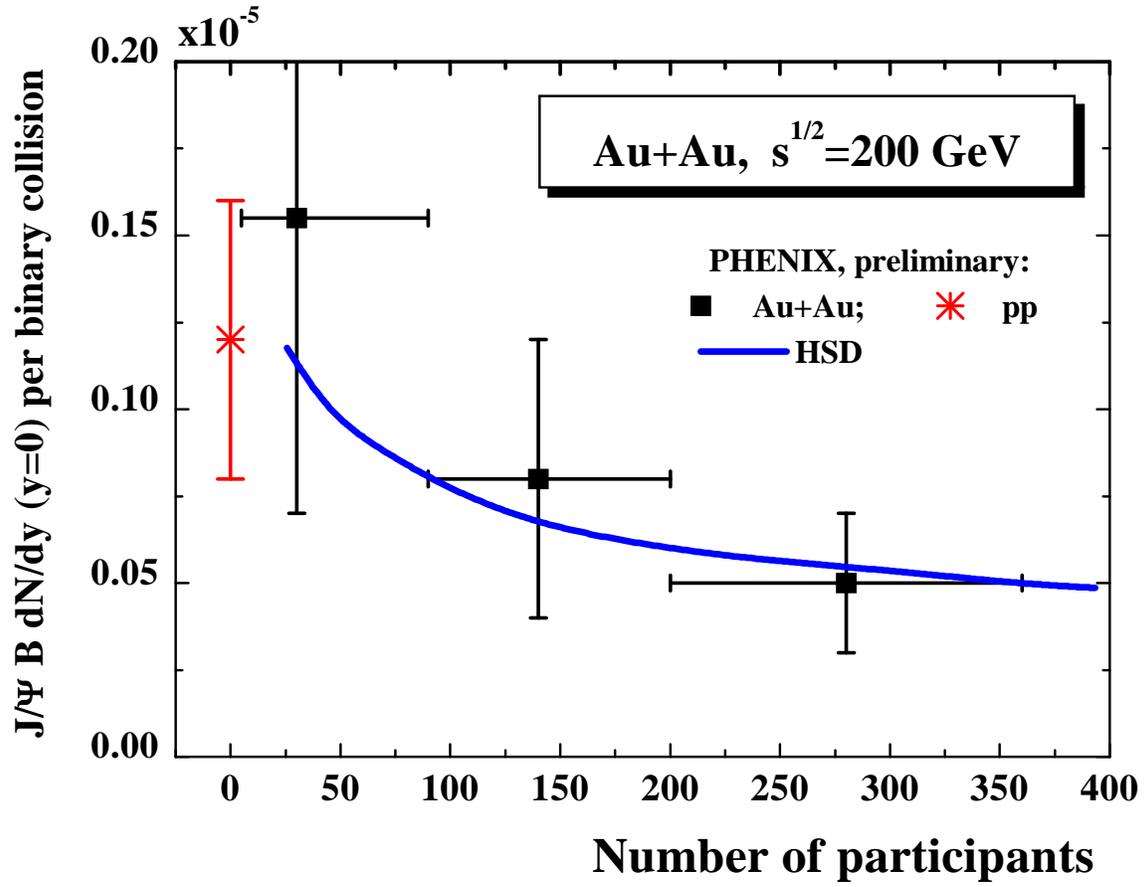,width=15cm}}
\vspace*{5mm}
\caption{The calculated $J/\Psi$ multiplicity per binary collision --
multiplied by the branching to dileptons --  as a function of the
number of participating nucleons $N_{part}$ in comparison to the
preliminary data from the PHENIX Collaboration \protect\cite{PHENIX}
for $Au+Au$ and $pp$ reactions.}
\label{bild14n} \end{figure}

\vspace*{-5mm}
The authors acknowledge inspiring discussions with M. Bleicher, C.
Greiner and A. P. Kostyuk. Furthermore, they like to thank J. J.
Gaardhoje for providing the experimental data of the BRAHMS
Collaboration in Fig. \ref{bild2}.


\clearpage

\clearpage

\clearpage

\clearpage

\clearpage

\clearpage

\clearpage

\clearpage

\clearpage

\clearpage

\clearpage

\clearpage

\clearpage

\clearpage

\end{document}